\PassOptionsToPackage{dvipsnames}{xcolor}
\documentclass[
english,
aps,
a4paper,
prd,
groupedaddress,
nofootinbib,
preprint,
floatfix
]{revtex4-1}

\usepackage{graphicx} 
\usepackage[utf8]{inputenc}
\usepackage{amsmath}
\usepackage{amssymb}
\usepackage{array}
\usepackage{graphicx}
\usepackage{gensymb}
\usepackage{float}
\usepackage{geometry}
\usepackage[myheadings]{fullpage}
\usepackage{multirow}
\usepackage{url}
\usepackage{hyperref}
\usepackage{cleveref}
\usepackage[normalem]{ulem}
\usepackage{listings}
\usepackage{xspace}

\usepackage[frozencache]{minted}

\usepackage{xcolor} 
\definecolor{LightGray}{gray}{0.9}
\usepackage{mathtools}
\newcommand{\floor}[1]{\lfloor #1 \rfloor}

\def\lsim{\raise0.3ex\hbox{$\;<$\kern-0.75em\raise-1.1ex\hbox{$\sim\;$}}}

\crefname{appendix}{Appendix}{Appendix}

\def\BSMArt{{\sc BSMArt}\xspace}
\def\SPheno{{\sc SPheno}\xspace}
\def\SARAH{{\sc SARAH}\xspace}

\def\smodels{{\sc SModelS}\xspace}

\newcounter{mylisting}
\newcommand{\listingcaption}[1]{%
  \refstepcounter{mylisting}%
  \noindent\textbf{Listing \themylisting:} #1%
}

\newcommand{\AddrLIP}{%
 LIP --- Laborat\'orio de Instrumenta\c{c}\~ao e F\'isica Experimental de Part\'iculas, 
 Departamento de F\'isica,  Escola de Ciências, Universidade do Minho, Campus de Gualtar,
 4701-057 Braga, Portugal}
 
\newcommand{\AddrWurz}{%
    Inst. f\"ur Theoretische Physik und Astrophysik,
    Julius-Maximilans-Universit\"at W\"urzburg \\
        Campus Hubland Nord, Emil-Hilb-Weg 22,
D-97074 W\"urzburg, Germany}

\newcommand{\AddrIPPP}{
    Institute for Particle Physics Phenomenology,
    Durham University,
    Durham DH1 3LE,
    United Kingdom}

\newcommand{\AddrLPTHE}{
    Laboratoire de Physique Th\'eorique 
    et Hautes \'Energies (LPTHE), 
   UMR 7589, 
   Sorbonne Universit\'e et CNRS, 
   4 place Jussieu, 75252 Paris Cedex 05, 
   France}

\newcommand{\AddrLiv}{Department of Mathematical Sciences,
University of Liverpool,
Liverpool,
L69 7ZL,
United Kingdom}

\begin{document}

\title{BSMArt 2: simpler and faster parameter space scans
}

\author{Fernando Abreu de Souza}\email{abreurocha@lip.pt}
\affiliation{\AddrLIP}
\author{Nuno Filipe Castro}\email{nuno.castro@fisica.uminho.pt}
\affiliation{\AddrLIP}
\author{Miguel Crispim Rom\~ao}\email{miguel.romao@durham.ac.uk}
\affiliation{\AddrIPPP}
\affiliation{\AddrLIP}
\author{Mark~D.~Goodsell}\email{goodsell@lpthe.jussieu.fr}
\affiliation{\AddrLPTHE}
\author{Farid Ibrahimov}\email{fibrh@liverpool.ac.uk}
\affiliation{\AddrLiv}
\author{Werner Porod}\email{porod@physik.uni-wuerzburg.de}
\affiliation{\AddrWurz}

\begin{abstract}
We present version 2 of BSMArt, a powerful yet lightweight scanning tool designed to simplify the exploration of parameter spaces of new physics models. Aside from architectural improvements, simpler installation and expanded documentation with examples, the new version includes additional tools and new machine learning and Monte Carlo scanning algorithms: Affine MC, Contour Finding, MLScanner, DLScanner, MLS, and CMA-ES. We showcase two variants of CMA-ES scans with physics applications relevant for soft lepton excesses at the LHC. We demonstrate that it is easily possible to find diverse and interesting parameter points for future testing.
\end{abstract}

\rightline{IPPP/26/43}

\maketitle

\section{Introduction}

In the absence of a preferred candidate framework to extend the Standard Model of particle physics it is necessary for phenomenologists explore many different possibilities and look for clues to decide between them. Testing a particular model then requires comparing its predictions against as many observables as possible. Facilitating the passage from model idea to observables is therefore a significant branch of research. 

Models can in principle be defined by some choices or fundamental constraints at different energy scales: either the couplings and masses can be specified from the ``bottom up'' just above the electroweak scale; or ``top down'' from some unification, string, symmetry breaking, etc, scale. The masses of the particles and their couplings can also be defined at either scale, either as parameters or as ``on shell'' quantities defined in terms of physical observables (e.g. the mass of the Z boson is precisely measured). Deriving the relationship between the parameters of a new model and the masses and couplings of the particles it predicts at observable energies, taking into account known observations, is the task of a spectrum generator: for example \SARAH \cite{Staub:2008uz, Staub:2013tta, Goodsell:2014bna, Goodsell:2017pdq} can create a bespoke code linked to the \SPheno \cite{Porod:2003um, Porod:2011nf} library (which is often just referred to as \SPheno code) or the related package {\sc FlexibleSUSY} \cite{Athron:2014yba,Athron:2017fvs,Athron:2022isz,Kotlarski:2026zmq} can create a spectrum generator linked to the {\sc SoftSUSY} \cite{Allanach:2001kg} library. But this is just the first step of a chain of tools: the outputs of these codes then need to be checked against measurements of the Higgs boson's mass and couplings; collider searches for new particles; searches for dark matter; flavour constraints; checks of stability of the vacuum; etc. There are different competing tools for each of these observations, and it is either complicated or tedious to chain them together to test all aspects of a model. 

This problem has arisen recently in the context of attempting to explain certain excesses in Large Hadron Collider (LHC) data. Four experimental analyses \cite{ATLAS:2019lng,ATLAS:2021moa,CMS:2021edw,ATLAS:2021kxv,CMS:2021far} have shown excesses of events above the backgrounds in channels either involving jets plus missing energy, or soft leptons plus missing energy. These excesses were shown to potentially have a common explanation \cite{Agin:2023yoq,Agin:2024yfs} which could be interpreted as electroweak-charged light particles. However, it remains an open question as to what model provides the best explanation: in the context of supersymmetry, they could be a higgsino-like or wino-bino like in the MSSM \cite{Baer:2023olq,Chakraborti:2024pdn,Martin:2024pxx,Constantin:2025bqp}, or a sector of the NMSSM \cite{Ellwanger:2024vvs,Araz:2025bww,Bagnaschi:2025ksk}, or possibly some other explanation could fit the data best. Even once a model has been chosen, finding the region of parameter space which provides the \emph{best fit} to the data is a technical challenge; in \cite{Agin:2025vgn} this was attempted in the simplified wino-bino scenario by combining collider searches, but in realistic models we need to scan over many parameters and compare to many additional constraints (dark matter, electroweak precision, etc) prior to the use of heavy machinery of collider recasting. In this paper we will use new techniques to do exactly that, solving the problem of finding relevant and interesting scenarios to test.

Since its original version \cite{Goodsell:2023iac,Faraggi:2023jzm}, \BSMArt has provided a lightweight and simple yet powerful way to quickly start parameter space scans of models using a wide set of codes relevant for dark matter and collider observables, while being flexible and sophisticated enough to run intense investigations. It could therefore streamline all aspects of investigating the properties of a general model from initial testing (``have I implemented the model correctly?''), to validation of a new precision computation (``how much does this correction change the Higgs mass?'') to complete determination of the valid or interesting parameter space. The primary motivation was to facilitate models implemented in \SARAH, for which scripts were provided to handle all the steps of the code generation and installation, and creation of templates for the scans. However, due to the simple definition of scans in terms of {\tt json} files, it is also not only possible but highly convenient to use \BSMArt without \SARAH or \SPheno or even any spectrum generation, not only for scans but also for tasks such as just writing Les Houches input files in batches based on some template. 

Other tools were or have become available for accomplishing similar tasks, with different motivations. {\sc GAMBIT} \cite{Martinez:2017lzg} is a heavyweight framework intended for large scans on computer clusters; its philosophy is to remove writing of intermediate files and build all codes into one monolithic program via ``backends.'' The output of the chain of codes is a global likelihood, explored by scanning algorithms such as {\sc MultiNest} \cite{Feroz:2007kg,Feroz:2008xx,Feroz:2013hea} or {\sc Diver} \cite{Martinez:2017lzg}.  Since it is written in {\tt c++} and is somewhat large/complicated to install, however, implementations of scanning algorithms employing machine learning (ML) techniques have been written independently in {\tt python}. Since the exploration of model parameter spaces is ubiquitous in HEP, recently there has been a proliferation of such tools. For example, {\tt xBit} \cite{Staub:2019xhl} (the precursor to \BSMArt);  \BSMArt version 1 \cite{Goodsell:2023iac,Faraggi:2023jzm}; {\sc EasyScanHEP} \cite{Shang:2023gfy}; {\sc MLScanner} \cite{Hammad:2022wpq} and {\sc DLScanner} \cite{Hammad:2024tzz}; {\sc Hep-aid} \cite{Diaz:2024sxg}; {\sc Jarvis-HEP} \cite{Guo:2026kfy}. An interesting tool called {\sc NMSSMScanner} to explore the NMSSM using a modified version of \BSMArt has also recently been announced \cite{Boto:2026bmz}.

A unique feature of \BSMArt is the scanning philosophy: there is no one-size-fits-all approach to scanning the parameter space of a model. Most development of scanning in HEP adopts likelihood-based sampling/inference techniques, because using a (log) likelihood function is the best way to combine multiple different measurements (as opposed to e.g. declaring that a point is excluded if it fails just one test);  see \cite{AbdusSalam:2020rdj} and references therein. Computing a global likelihood for each point then allows the use of powerful sampling techniques that have been developed for other fields: Markov-Chain Monte Carlos (MCMCs) and related algorithms such as {\sc MultiNest} \cite{Feroz:2007kg,Feroz:2008xx,Feroz:2013hea}; differential evolution such as {\sc Diver} \cite{Martinez:2017lzg}; and modern tools such as simulation-based inference or Bayesian optimisation. Already this comprises a large number of algorithms and \BSMArt aims to make the use or implementation of any of these as simple as possible by abstracting the (parallel) running of tools from the task of choosing which points to test. Its implementation in {\tt python} makes the use of ML techniques straightforward.

However, the primacy of the likelihood paradigm obscures other needs for scanning in HEP.
In initial testing of the implementation of a model, the individual outputs are very important (e.g. checking the Higgs boson or other particles' masses) and we want simple two-dimensional plots of these, often as a linear grid. As emphasised in \cite{Goodsell:2022beo}, where an Active Learning (AL) scan was developed, sometimes the object of interest is the \emph{boundary} of the exclusion region, and sampling many valid points in high-likelihood regions is not desirable, especially if the evaluation time of the tools used is long. In addition, sometimes our tool may only give a binary ``valid/not-valid'' response rather than a continuous function. Other uses include trying to find parameter points with a particular property, such as large or small mass splittings between two states, or finding the heaviest possible dark matter candidate allowed by the model \cite{Goodsell:2020rfu,Goodsell:2022beo}: for these we need to bias our search, rather than only using the likelihood corresponding to the compatibility of the observables with experimental data. There is also an argument that in scans using ML techniques, by working \emph{only} with a likelihood -- as opposed to also feeding observables or other quantities to the network -- that we are not making full use of the available information. \BSMArt was designed to accommodate all of these, by allowing both likelihood-based scans and other possibilities; but also allowing flexible definitions of likelihoods by the user with functions for biases. The aim is to have as large a menu of different scans as possible.

The initial set of included scans was powerful enough for many purposes: basic choices such as  Grid and Random scans; convenience scans for reading parameters from a {\tt csv} file, or using a predetermined set of Les Houches input files; a simple (yet parallel) MCMC for general use;  {\sc MultiNest} and {\sc Diver} scans using their external libraries (notably these scans are those used by {\sc GAMBIT}; and a novel Active Learning scan \cite{Goodsell:2022beo} mentioned above. However, this has been an area of active development and many new algorithms have been developed, including \cite{Goodman:2010dyf,lukas_heinrich_2018_1634428,deSouza:2022uhk,Romao:2024gjx,deSouza:2025uxb,deSouza:2025bpl,deSouza:2026sta}. In particular, aside from {\sc MultiNest} and {\sc Diver} which rely on external libraries, the original scans consisted of algorithms that scaled poorly with increasing numbers of model parameters and were therefore not suitable for complicated models. In this paper we present the implementation in \BSMArt of several algorithms to fill the gap. In particular, we implement a native affine MCMC scan based on the Goodman and Weare algorithm \cite{Goodman:2010dyf}; a CMA-ES algorithm \cite{deSouza:2022uhk} and an algorithm using novelty detection with CMA-ES based on \cite{Romao:2024gjx}. We also describe two contour-finding scans: a two-dimensional contour-finder; and a contour-finder using Gaussian Processes based on/extended from the {\tt excursion} \cite{lukas_heinrich_2018_1634428} package.

Since the initial version, the structure of the code and its documentation was modified to allow it to be installed as a {\tt python} package. Since \BSMArt is written by authors of the \SARAH package, there is a guarantee of compatibility for those tools and related packages. The implementation of scans as independent scripts with the actual running of the tools abstracted away, made even easier now that \BSMArt or its components can be imported, means that developers can implement their own algorithms using the \BSMArt toolchain without having to reinvent the wheel and confident that their code will be supported in the future. To accompany the new version we therefore have created a {\tt github} repository\footnote{Available at \url{https://github.com/BSMArt-HEP/examples}} for sharing community scans and tools.

The new installation instructions are described in section \ref{SEC:INSTALLATION}.
As part of the development of the additional scans, several new features were added, which we describe in section \ref{SEC:FEATURES}. 
Interfaces to several new tools have been included, which we describe in section \ref{SEC:TOOLS}. We describe the new scans in sections \ref{SEC:PORTED} and \ref{SEC:CMAES}. We then finally utilise the power of the new code and algorithms for physics applications  relevant for searches for soft leptons at the LHC in section \ref{SEC:PHYSICS}. 

\section{Installation and model setup}
\label{SEC:INSTALLATION}

Version 2 of \BSMArt can now be installed as a python package:
\begin{minted}
[
frame=single,
framesep=2mm,
baselinestretch=1.2,
bgcolor=White,
fontsize=\footnotesize,
mathescape=true,
linenos=false,
]
{bash}
python3 -m pip install --user bsmart
\end{minted}
Otherwise the files can be downloaded from the website and installed locally via tarball or wheel. 
In both cases we highly recommend the use of a virtual environment, e.g. 
\begin{minted}
[
frame=single,
framesep=2mm,
baselinestretch=1.2,
bgcolor=White,
fontsize=\footnotesize,
mathescape=true,
linenos=false,
]
{bash}
python3 -m venv BSMArt_venv
source BSMArt_venv/bin/activate
python3 -m pip install bsmart
\end{minted}

The code is now hosted both on the original website:
\begin{center}\url{https://goodsell.pages.in2p3.fr/bsmart/}\end{center}
where we  maintain both the legacy version {\tt v1.7} and latest release;
and version {\tt 2.0.0} onwards is mirrored on github:
\begin{center}\url{https://github.com/BSMArt-HEP/core/}\end{center}
with online documentation at:
\begin{center}\url{https://bsmart-hep.github.io/core/index.html}\end{center}

{\tt tcsh} users should also type {\tt rehash} after installation to update paths; the installation will create several runnable scripts. For a fresh installation, it is recommended to use {\tt BSMArt-InstallHEPTools} to install the various HEP codes (\SARAH, \SPheno, {\sc MicrOMEGAs} etc) desired. In the process it will create a file {\tt HEPtoolpaths.json} containing the paths of the necessary tools. In the new version, this file will preferentially be stored in a subdirectory of the virtual environment, or the user's home directory, unless the {\tt --local} flag overrides this to store it in the working directory.

To begin using \BSMArt, there is a quick start script:
\begin{minted}
[
frame=single,
framesep=2mm,
baselinestretch=1.2,
bgcolor=White,
fontsize=\footnotesize,
mathescape=true,
linenos=false,
]
{bash}
BSMArt-QuickStart
\end{minted}
This will build the code for the MSSM and create two simple example scans, configured for all the tools installed. 

For general use, the command
\begin{minted}
[
frame=single,
framesep=2mm,
baselinestretch=1.2,
bgcolor=White,
fontsize=\footnotesize,
mathescape=true,
linenos=false,
]
{bash}
BSMArt-PrepareModel --All MODELNAME 
\end{minted}
will build and install the outputs from \SARAH for the corresponding model, configured for all installed tools (individual tools can be specified instead of the {\tt --All} flag). As an alternative to {\tt SPheno}, {\sc FlexibleSUSY} can be specified as the spectrum generator and the code will be generated and compiled accordingly. 

The above scripts will also create a working directory of the form {\tt BSMArt-MODELNAME} into which template Les Houches files, a template {\tt json} file for a scan (configured with the paths for the relevant codes) and other files required for different tools (e.g. {\tt parameters.ini} for {\sc SModelS}) are placed. By navigating to this directory and editing the templates, a \BSMArt scan can be launched via 
\begin{minted}
[
frame=single,
framesep=2mm,
baselinestretch=1.2,
bgcolor=White,
fontsize=\footnotesize,
mathescape=true,
linenos=false,
]
{console}
cd BSMArt_MODELNAME
BSMArt Template_MODELNAME.json 
\end{minted}
For debugging information the {\tt --debug} flag can be added. Typing {\tt BSMArt} without specifying a json input file will instead give information about available scans and tools; typing {\tt BSMArt <tool name>} or {\tt BSMArt <scan name>} will give information about the options/necessary parameters for the corresponding tool or scan. The specifications for the input {\tt json} file are given in the original \BSMArt manual \cite{Goodsell:2023iac} and in the new online documentation\footnote{\url{https://bsmart-hep.github.io/core/index.html}}; we remind the reader that the template Les Houches files should be edited so that any value that should be varied in a scan (by being replaced by a variable or a function of a variable) is just given as a text value, e.g.:
\begin{minted}
[
frame=single,
framesep=2mm,
baselinestretch=1.2,
bgcolor=White,
fontsize=\footnotesize,
mathescape=true,
linenos=false,
]
{text}
BLOCK MINPAR # Input parameters
 1 m0 # m0
 2 m12 # m12
 3 1.0000000E+01 # TanBeta
 4 1.0000000E+00 # SignumMu
 5 - 2.0000000E+03 # Azero
\end{minted}

\subsection{Examples}

The command
\begin{minted}
[
frame=single,
framesep=2mm,
baselinestretch=1.2,
bgcolor=White,
fontsize=\footnotesize,
mathescape=true,
linenos=false,
]
{bash}
BSMArt-BuildExamples 
\end{minted}
will build the code and configure scans for a large set of examples bundled with the package.

However, a new feature introduced with version 2 is the community repository at \url{https://github.com/BSMArt-HEP/examples}. It is intended as a store of example scans and tools with the hope that users will contribute their own. Configuring the scans for use with the tools installed on uses' computers is now straightforward; once the code for the model has been build via the {\tt BSMArt-PrepareModel} script, a template can be configure via
\begin{minted}
[
frame=single,
framesep=2mm,
baselinestretch=1.2,
bgcolor=White,
fontsize=\footnotesize,
mathescape=true,
linenos=false,
]
{bash}
BSMArt-PrepareBSMArt <modelname> --template=<template-path>
\end{minted}
This will supply the paths for the codes and include relevant data (such as the particle codes for the Higgs bosons necessary for {\sc HiggsTools}) for the template scan, which can then be run using the {\tt BSMArt} command. 

\subsection{Code outputs}

Since \BSMArt attempts to fulfil many different needs, there are many different possible outputs. When a scan starts running, it creates a subdirectory of the working directory with the name specified by the {\tt "RunName"} option in the {\tt "Setup"} block of the {\tt json} input. Inside this directory will be:
\begin{itemize}
    \item A {\tt references.bib} containing bibtex information for the tools and scans used in that run, intended to aid paper writing. 
    \item A {\tt Plots} subdirectory. \BSMArt provides the ability to automatically generate simple plots based on the collected data. In version 2, each scan will now also automatically generate a corner plot to show the distribution of sampled points amongst the variables. 
    \item A {\tt Spectrum\_Files} subdirectory, containing information gathered about the parameter points scanned, detailed below.
    \item Possible other subdirectories generated by different scans (e.g. some scans produce a {\tt Results} subdirectory). 
\end{itemize}

The outputs in the {\tt Spectrum\_Files} subdirectory include as different options in the {\tt "Setup"} dictionary of the {\tt json} input:
\begin{itemize}
\item To collect all the outputs of the different codes into a separate directory for each parameter space point. This is obviously suitable only for reasonably small scans, and selected via the option {\tt "StoreEverything":"True"}.
\item A separate Les Houches file for each point, or all such files concatenated into one large file. For most tools this is all that is necessary, as within \BSMArt they write their outputs into a ``spectrum file'' that is passed from one tool to the next within the temporary working directory. However, keeping these files still takes significant disk space, so is not suitable for large or fast scans. If  {\tt "StoreAllPoints":"True"} is selected then the spectrum files are stored; they are stored as separate files if {\tt "StoreSeparateFiles"} is also true (by default it is false). . 
\item {\tt csv} or tabbed outputs of ``variables''  and ``observables,'' plus any ``result'' value determined by the particular scan. This is the typical usage: any interesting information is extracted as an ``observable'' after the running of each code, and that is all that we need to keep. In this way, easily human or machine readable comma or tabbed separated values files can be stored which are relatively economical. If the user decides afterwards that they have missed some quantities, rather than rerunning a new scan it is possible to just run selected points through the toolchain using a {\tt read\_csv} or {\tt read\_dir} scan. {\tt csv} output is selected via {\tt "csv":"True"} and tabbed output via {\tt "Short":"True"}.
\end{itemize}

\section{New features}
\label{SEC:FEATURES}

\subsection{Functions}

At the core of \BSMArt's running is the idea that editing input files for a scan should be as easy as possible. As described above, some values (e.g. options in the {\tt SPHENOOPTIONS} block) should be fixed, while others are varied, either being variables of a scan or functions thereof. In the new version, it is now straightforward to define new observables that are functions of scan variables or observables. Such quantities may be interesting as observables in themselves, or we may want to use them for validity criteria or as part of a likelihood computation. The syntax for observables is now
\begin{minted}
[
frame=single,
framesep=2mm,
baselinestretch=1.2,
bgcolor=White,
fontsize=\footnotesize,
mathescape=true,
linenos=false,
]
{json}
"SLHAobsname" : { "SLHA": ["BLOCKNAME",[BLOCKNUMBERS]], ...},
"FUNCTIONobsname" : { "FUNCTION": "function of variables and named observables", ...},
\end{minted}
The options in ellipsis including {\tt "SCALING"} for use in likelihoods, and {\tt "VALID"} for specifying a validity condition. Examples of usage are given in section \ref{SEC:PHYSICS} and in the examples repository. 

Similarly, we may want \emph{variables} to be derived as functions from other variables. To this end, the {\tt json} file may now have a new section called {\tt "Derived Variables"} which is a dictionary of new variable names to a string containing the value, e.g. 
\begin{minted}
[
frame=single,
framesep=2mm,
baselinestretch=1.2,
bgcolor=White,
fontsize=\footnotesize,
mathescape=true,
linenos=false,
]
{json}
{
   "Variables": { "m0" : { "RANGE": [200,4000]}},
   "Derived Variables": { "m02": "m0**2"}
}
\end{minted}

\subsection{Datapoint objects}
\label{sec:datapoint}

Since the tools are executed in order (until/unless a validity condition is failed) the observables collected and derived by each tool are available for the subsequent ones. This means that purely pythonic tools can be created by the user to perform arbitrary manipulations. In order to make this as simple as possible, the data collected by the chain of tools is now stored in a {\tt DataPoint} class. The variables are accessible via the method {\tt get\_var\_dict()} and all variables and observables via {\tt get\_all\_dict()}

For example, in testing scan algorithms we may want to create a tool that generates a fake likelihood based on input variables without having to read/write Les Houches input files. We can create a tool placed in a subdirectory {\tt Tools} of the current directory where the scan is launched, and, if the tool is called {\tt mytool} in the {\tt "Codes"} section of the {\tt json} file, then we create a python script {\tt Tools/mytool.py}. Here is an example of such a script that outputs a Gaussian likelihood with an invalid region in a rectangular hole:
\begin{minted}
[
frame=single,
framesep=2mm,
baselinestretch=1.2,
bgcolor=White,
fontsize=\footnotesize,
mathescape=true,
linenos=false,
]
{python}

from bsmart import debug
from bsmart.HEPRun import HepTool, DataPoint

import numpy as np
import math
class NewTool(HepTool):
    def __init__(self, name, settings,global_settings=None):
        HepTool.__init__(self, name, settings,global_settings)
        
    def run(self, spc_file, temp_dir, log,data_point):    
        vars = data_point.get_var_dict()
        m0=vars['m0']
        m12=vars['m12']

        # Fake a 'dead' region:

        if m0 < 1400 and m0 > 1100 and m12 < 1500 and m12 > 1200:
            raise
        
        # let's scale things
        x0=m0/2500.0
        x1=m12/2500.0
        a0=2000/2500.0
        a1=1100/2500.0
         
        
        # double gaussian loglikelihood
        res=np.exp(-((x0-a0)**2 +(x1-a1)**2)/0.01)
        data_point.obs_dict['fres'] = res
\end{minted}
The observable created by this tool, called {\tt fres}, is then available to the code; e.g. a {\tt json} file to sample from this using the new AffineMC scan might look like:
\begin{minted}
[
frame=single,
framesep=2mm,
baselinestretch=1.2,
bgcolor=White,
fontsize=\footnotesize,
mathescape=true,
linenos=false,
]
{json}
{
	"Codes": {
		"mytool": {
			"InputFile": "fake.in",
			"Run": "True",
			"Observables": {
				"lres": {
					"FUNCTION": "fres",
					"SCALING": "USER"
				}
			}
		}
	},
	"Setup": {
		"RunName": "compareEMCEE",
		"Type": "AffineMC",
		"Cores": 4,
		"StoreEverything": "False",
		"Skip SLHA Files": "True",
		"StoreAllPoints": "False",
	    "Store In Memory": "True",
	    "Store Invalid": "True",
		"StoreSeparateFiles": "False",
		"csv": "True",
		"Merge Results": "True",
		"Output File": "Fake_Output",
		"Spectrum File": "fake.in",
		"Steps": 100
	},
	"Networks": {},
	"Variables": {
		"m0": {
			"RANGE": [
				200,
				2500
			],
			"VARIANCE": 10
		},
		"m12": {
			"RANGE": [
				200,
				2500
			],
			"VARIANCE": 10
		}
	},
	"Observables": {},
	"Plotting": {
		"Strategy": "csv",
		"Plots": {}
        }
}
\end{minted}

\subsection{Improvements to the likelihood computation}

In \BSMArt, the computation of likelihoods is not automatically performed, but depending on the scan the {\tt BSMlikelihood} subpackage can be used. This means that user-defined scans are at complete liberty to create their own likelihood routines; but the bundled scans now all use the inbuilt ones to unify the notation. 
Each observable can be associated with a likelihood if it has an attribute {\tt "SCALING" } that is not {\tt "OFF"}; the global likelihood is then the product of all likelihoods. In practice, it is now the \emph{scan} which decides whether it will actually work with a likelihood, log-likelihood or negative log likelihood, and the routines will compute the corresponding quantity (e.g. if the scan requires a log-likelihood it will directly compute that and not first take a likelihood); this means the {\tt "LogLike"}  option in previous versions of \BSMArt is largely ignored, except for certain scans (where the option is indicated) where it now instructs the scan to use a log likelihood. 

In the latest version, likelihoods are therefore specified in a unified way, with {\tt SCALING} options which include:
\begin{itemize}
\item {\tt "OFF"} for observables to be ignored for the likelihood, but may otherwise be interesting for plots, etc.
\item {\tt "GAUSS"} for a gaussian likelihood; this is the default choice if no other option is taken and a {\tt "MEAN} and {\tt "VARIANCE"} are provided. 
\item {\tt "UPPER"} and {\tt "LOWER"} for a function that cuts off softly at an upper or lower bound (via a sigmoid function, with the softness controlled by a {\tt "VARIANCE"} setting)
\item {\tt "BIAS"} for a power-law likelihood. If the observable value is $x$, the {\tt "MEAN
} is $m$ and the {\tt "VARIANCE"} is $s$, then the bias likelihood is $(x/m)^s$. This is useful for cases where we want to push the scan in a certain direction.
\item User supplied values. These are the most powerful choices, either to use a likelihood computed by a specific code (e.g. the $p$-value from {\sc MicrOMEGAs} or {\tt HiggsTools}); or a value computed by a user-defined tool (as in the fake example scan above); or as the value computed using the new facility to define functions. Since different codes return different objects, such as $p$-values or log-likelihoods, there are options:
\begin{enumerate}
\item {\tt "USER"} when the observable is already a likelihood/$p$-value
\item {\tt "SIGUSER"} to apply a sigmoid function to a user-supplied value. This allows a soft cutoff, to safely handle extreme values.
\item {\tt "EXPUSER"} when the supplied value is a log-likelihood so we would need to exponentiate it to obtain a likelihood. 
\item {\tt "MINUSEXPUSER"} when the supplied value is a negative-log-likelihood so we would need to exponentiate the negative of it to obtain a likelihood.
\end{enumerate}
\item {\tt "CFUNCTION"} is a special function that gives zero within a specified range, and increases linearly outside of it. This is intended for use with the new {\tt CMAES\_ND} scan described in the following sections. 
\end{itemize}

In addition, the likelihood routines have been adapted to allow for the case where only a subset of the observables have actually been calculated. It is often desirable to implement validity conditions, whereby the toolchain stops if preliminary conditions are not met; the simplest case is when no spectrum file has been produced by \SPheno, where we have no information and the likelihood should be minimal. But perhaps we do not want to compute the relic density if we already know that the wrong particle is the lightest state in the spectrum, or if the Higgs boson mass is not in the correct range. In such cases, we \emph{do} have some information and may be able to use it to direct the scan, if the scan uses a log-likelihood or negative-log-likelihood: in such cases we set the individual contributions of all the missing observables that have not been computed to a fixed large-magnitude value. In this way, we can save significant computing time (and avoid problems running later codes with potentially unphysical points). More details and examples of using this feature are given in section \ref{SEC:CMAES}, which requires the following setup options:
\begin{minted}
[
frame=single,
framesep=2mm,
baselinestretch=1.2,
bgcolor=White,
fontsize=\footnotesize,
mathescape=true,
linenos=false,
]
{json}
{
    "Setup": {
        "Store Invalid": "True",
        "Store In Memory": "True"
    }
}
\end{minted}
These direct \BSMArt to not discard invalid points, and to keep a copy in memory of the collected variables and observables of each batch (which are then accessible by the scan). If either of these is false, then points failing any validity condition will have maximal/minimal loss for all observables.

\subsubsection{Priors}

In general, algorithms such as MCMCs are designed to asymptote to sampling from the posterior distribution. Bayes' theorem tells us that 
\begin{align}
P(x|D) =& \frac{P(D|x) \pi(x)}{P(D)} \equiv \frac{P(D|x) \pi(x)}{Z},
\end{align}
where $P(D|x) $ is the probability of observing data $D$ given variables $x$, $\pi (x)$ is the \emph{prior probability} of variables $x$, $P(D) \leftrightarrow Z$ is the probability of data $D$, which is just a normalisation. and $P(x| D)$ is the object of interest, the probability of variables $x$ given data $D$. We determine $Z$ as normalisation:
\begin{align}
    \int dx P(x|D) = 1 \longrightarrow Z = \int dx P(D|x) \pi(x).
\end{align}
The standard method of implementing prior probabilities $\pi(x)$ into likelihood-based scans is to sample from a unit hypercube and adjust the mapping from that hypercube to the variable search space. If the variable is allowed to have values between values $[v_{\rm min}, v_{\rm max}]$ then a linear mapping from $[0,1]$ to this obviously corresponds to a uniform prior with normalisation $1/|v_{\rm max} - v_{\rm min}|$. But if we suppose that our algorithm samples points with distribution $P(D| x)$, then it returns $P(D|x) dx$ points in a region around $x$. Making a change of variables we have
\begin{align}
 P(D|x)  dx = P(D| x) \left| \frac{dx}{dy}\right| dy
\end{align}
so since we have $\pi(x)= \mathrm{const},$ if we choose $y$ such that 
\begin{align}
\pi(y) = \left| \frac{dx}{dy}\right| 
\end{align}
then drawing samples from $x$, transforming to variable $y$ and computing $P(D| y)$ will be the same as drawing samples from $P(D| y) $ with prior $\pi(y)$.

In \BSMArt the use of priors is therefore possible for every scan using the scaler functions from {\tt bsmart.BSMutil} to rescale from a hypercube; this includes {\tt AffineMC, CMAES, MultiNest, Diver, MLS, MLScanner, DLScanner} algorithms. The default is a uniform prior in the variable range; other choices are activated by including {\tt "PRIOR": <TYPE>} in the variable definition, along with {\tt "RANGE": [vmin,vmax]} or a {\tt MEAN} and \texttt{SIGMA} setting, e.g. 
\begin{minted}
[
frame=single,
framesep=2mm,
baselinestretch=1.2,
bgcolor=White,
fontsize=\footnotesize,
mathescape=true,
linenos=false,
]
{json}
{
    "Variables": {
        "m0": {"PRIOR": "GAUSS", "MEAN": 1000, "SIGMA": 100},
        "m12": {"PRIOR": "LOG", "RANGE": [100,1000]}
    }
}
\end{minted}
The currently supported priors are \texttt{UNIFORM}; \texttt{GAUSS}; \texttt{SYMLINEAR}, \texttt{INTERVAL}; \texttt{OPEN}; \texttt{LOG}; \texttt{LOGNORMAL}; \texttt{EXP}; \texttt{SIN}; \texttt{CAUCHY}; \texttt{NONE} (no scaling is applied) and \texttt{FIXED} (all points are mapped to a single value).

\subsection{Automatic corner plots}

\begin{figure}[!h]\centering
\includegraphics[width=0.45\textwidth]{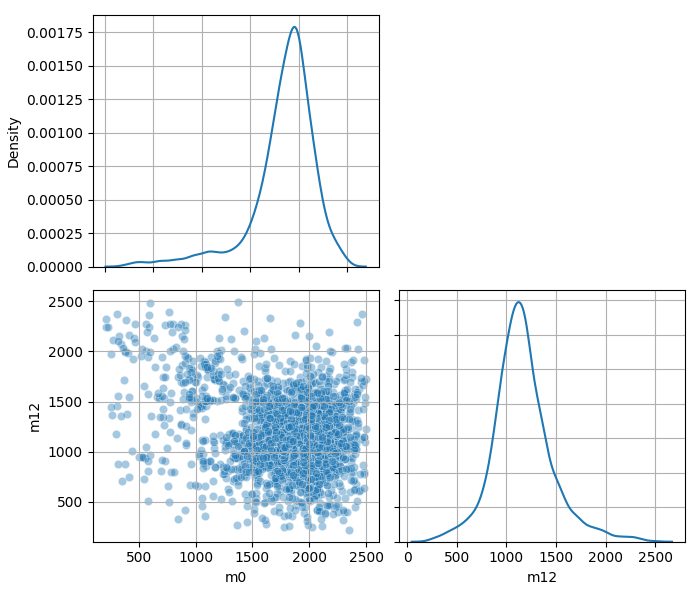} \includegraphics[width=0.45\textwidth]{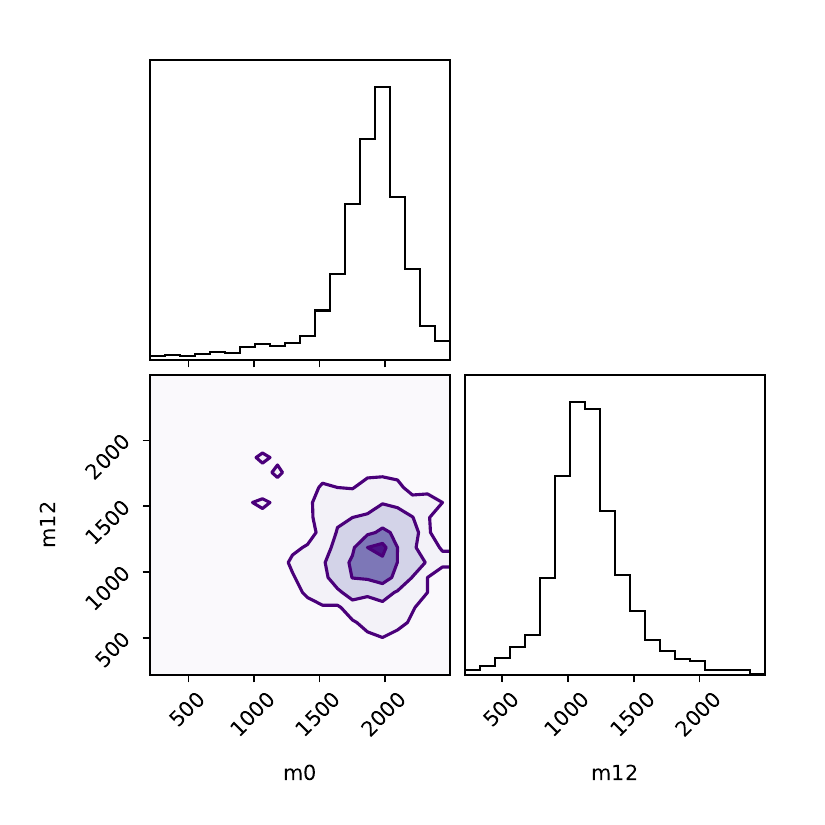} 
\caption{\label{FIG:examplecorners} Example corner plots for an Affine MCMC scan of a toy gaussian likelihood with a `hole'. Left: scatter plots; right: contours using the {\tt corner} package.}
\end{figure}

The plotting capabilities of \BSMArt have been extended so that, when plots are produced, an additional two \emph{corner plots} are now created automatically. These are used to show the correlations between the variables of the scan, with one plot for each pair of variables and an additional one showing the distribution of values for each variable, arranged in a pyramid shame. The first corner plot is a scatter plot; the second shows contours generated using the {\tt corner} package. The scripts to produce these are also provided, so that the user can modify them and adjust to use in papers as desired. We give an example in figure \ref{FIG:examplecorners} of an \texttt{AffineMC} scan of a likelihood using the fake tool \texttt{mytool} from section \ref{sec:datapoint}, which gives a gaussian likelihood with a `hole' in the parameter space where the points are declared invalid. The hole is visible in the scatter plot whereas the contours show the most likely region.

\subsection{Other new options}

Other new options in the {\tt Setup} dictionary include:
\begin{itemize}
\item {\tt "SLHA\_Output":"False"} tell \BSMArt to not look for output SLHA files; implicitly any observables will be collected internally in the tool running. This is useful for pythonic tools. 
    \item {\tt "Skip SLHA Files" :"True"} tells \BSMArt to skip both reading and writing of Les Houches files. This is relevant for pythonic tools, and especially useful for testing scans.
    \item {\tt "Precision":int} indicates that the output {\tt csv} or tabulated values files should be written with the number of digits precision indicated. This is useful for saving disk space/speeding up file input/output, and makes the results more easily human-readable\footnote{We thank Asesh Datta for suggesting this option.}. 
\end{itemize}

The {\tt Fitting} section of the {\tt json} input has also been extending to allow overloading of options during the optimisation phase. For example, it may be desirable to turn off the (slow) computation of particle decays when trying to fit a quartic coupling value to give the correct Higgs boson mass. This example is accomplished via:
\begin{minted}
[
frame=single,
framesep=2mm,
baselinestretch=1.2,
bgcolor=White,
fontsize=\footnotesize,
mathescape=true,
linenos=false,
]
{json}
{
    "Fitting": {
        "Variables": { "lamH" : {"RANGE": [-0.2,0.7]} },
        "Observables": {"mh" :{"TARGET": 125,"VARIANCE":0.5}},
        "Options" : {"eps": 0.1, "ftol" : 1.0, "disp" : "False", 
            "Overloads": [["SPHENOINPUT",[11],0]]}
    }

}
\end{minted}
Note that the observables used for fitting \emph{must} have already been declared as an observable from some tool or globally (if it is in one tool, then the optimisation routines will only run the relevant tools during the fitting phase) whereas the \emph{variables} must \emph{not} be declared elsewhere (since they are not scanned over, but fitted).

\section{Tools}
\label{SEC:TOOLS}

Since the original version \cite{Goodsell:2023iac} several new tools have been added.  The current list comprises:
\begin{itemize}
\item {\tt SPheno} meaning either \SPheno or \SARAH-generated fortran code compiled against the \SPheno library. 
\item {\tt FlexibleSUSY} \cite{Athron:2014yba,Athron:2017fvs,Athron:2022isz,Kotlarski:2026zmq} meaning the code produced by {\sc FlexibleSUSY} linked against the {\sc SoftSUSY} library, is a new option to run as an alternative to {\tt SPheno}. 
\item {\tt HiggsTools} \cite{Bahl:2022igd} to test Higgs boson properties and LHC searches for Higgs-like bosons was documented in the original version of \BSMArt. {\tt HiggsBounds} and {\tt HiggsSignals} are still available as options but deprecated. 
\item {\tt MicrOmegas} \cite{Belanger:2010gh,Alguero:2023zol} for computing the dark matter relic density and direct detection quantities was documented in the original version of \BSMArt. The code was been updated in line with newer version of {\sc MicrOMEGAs} which allows for multiple dark matter candidates. As described in \cite{Faraggi:2023jzm}, the interface code will also compute all pair production cross-sections of new particles, and also single-particle production for $Z'$ gauge bosons, provided the {\tt "cross-sections":"True"} option is set. These are required inputs for {\sc SModelS} (see below).
\item {\tt VevaciousPlusPlus} \cite{Camargo-Molina:2013qva}\footnote{The original version of {\sc Vevacious} no longer runs; while there is no paper for the {\tt c++} version, the code is available online: \url{https://github.com/JoseEliel/VevaciousPlusPlus/} and can be installed by \BSMArt.} checks for stability of the vacuum, as documented in the original \BSMArt paper.
\item {\tt flavio} \cite{Straub:2018kue} computes flavour constraints, as as documented in the original \BSMArt paper. 
\item {\tt MadGraph}  for event generation and cross-section computation via {\sc MG5\_aMC} \cite{Alwall:2014hca}. The running has been improved since the original \BSMArt version. 
\item {\tt MadGraphHackAnalysis} and {\tt HackAnalysis\_LO} run {\sc HackAnalysis} either via {\sc MadGraph} or with event generation in {\sc pythia} \cite{Sjostrand:2014zea,Bierlich:2022pfr}. These interfaces are documented in \cite{Goodsell:2024aig}. 
\item {\tt SModelS} \cite{Kraml:2013mwa, Ambrogi:2018ujg, Alguero:2020grj, Alguero:2021dig, Altakach:2024jwk,Constantin:2025bqp} which rapidly computes exclusion limits based on simplified topologies for LHC processes. It uses the particle production cross-sections computed by {\sc MicrOMEGAs} above together with the decay tables in the Les Houches file from \SPheno, and a {\tt QNUMBERS} file generated by \BSMArt from \SARAH. The interface was described in \cite{Faraggi:2023jzm}.
\item {\tt anyBSM} \cite{Bahl:2023eau,Bahl:2026nsu,anyHH} for computing predictions for, inter alia, triple Higgs couplings and di-higgs production. 
\item {\tt ZPEED} \cite{Kahlhoefer:2019vhz} computes limits on $Z'$ bosons decaying to lepton pairs; the interface was described in \cite{Faraggi:2023jzm}.
\item {\tt ZPrime} \cite{Alvarez:2020yim,Lozano:2021zbu} computes limits on $Z'$ bosons; the interface was described in \cite{Faraggi:2023jzm}. 
\item {\tt Resummino} \cite{Fuks:2013vua} computes higher-order production cross-sections for certain supersymmetric particles, and was employed in \cite{Agin:2024yfs,Agin:2025vgn}.
\item {\tt MadAnalysisExpert} and {\tt MadAnalysisAllPAD} provide interfaces to {\sc MadAnalysis} \cite{Conte:2012fm, Conte:2014zja, Conte:2018vmg} expert mode and Public Analysis Database (PAD, for running banks of existing analyses) respectively. 
\end{itemize}

In addition, we provide two virtual tools which are pure python scripts:
\begin{itemize}
\item {\tt pass\_tool} which does nothing\footnote{This has been renamed from \texttt{pass} in previous versions of \BSMArt, to avoid coincidence with a \texttt{python} keyword.}. It serves as a placeholder to allow the user to define observables and/or apply validity conditions. It is also useful as the initial tool in cases where there is no spectrum generator, where it is only desired to write the input Les Houches files (based on the choices of parameters) and keep them. 
\item {\tt toy\_targets} provides several different standard likelihood functions of few variables commonly used to test scans: Rosenbrock, Rastrigin, Ackley, Himmelblau, Easom. These will treat the input variables (whatever their names!) as an ordered list and compute the relevant function based on the first one or two of these. In this way, new scan algorithms can be easily tested without even needing to write any input/output files to disk.  
\end{itemize}

\section{Ported scans}
\label{SEC:PORTED}

Since one aim of \BSMArt is to provide a wide menu of scanning algorithms, in version 2 we ported several algorithms from elsewhere. We describe these below. 

\subsection{Affine MC}

Traditional MCMC scans for HEP implement the Metropolis-Hastings algorithm for choosing the next point in a chain: they propose a next point and then accept it with a probability that depends on a ratio of the likelihoods of the current and next point. To select which point to test, typically a gaussian step is used; in the MCMC scan in \BSMArt the variable values for the next point are chosen randomly from a gaussian distribution of mean equal to the current point and specified variances. In this way, the user must specify the desired variances and ranges for the variables in order to specify the scan. In order to have parallel execution, the code creates an independent chain for each processor core, which produce statistically independent results but at the expense of slower convergence than an algorithm that could work in batches. 

Such algorithms are the workhorse of simple HEP explorations because they are robust and easy to set up, while being capable of converging on regions of high likelihood. However, they are very inefficient when the number of variables is large, because the algorithm for generating the next step gains no information about which variables are most sensitive (the step size is not adapted). An algorithm that respects detailed balance but solves the problem of batch execution and high dimensions was proposed by Goodman and Wear \cite{Goodman:2010dyf}: an \emph{affine} MCMC. In this algorithm, a population of walkers is maintained; the population size should be a linear multiple of the number of dimensions, and the proposal for the next point is made by considering affine steps between pairs of points. In this way, the population adapts to the shape of the likelihood distribution and is therefore more efficient in adapting to higher dimensions.  

While a pythonic package implementing the Goodman and Wear algorithm called {\tt emcee} exists\footnote{See \url{https://emcee.readthedocs.io}.}, implementing an interface to it in \BSMArt was complicated because it insists on taking control of the parallelisation. This requires a separate running script, {\tt BSMArt\_emcee}. It also has the disadvantage that it does not store values for the observables in the output chains. We therefore implemented a native version of the algorithm from scratch, available in \BSMArt as an {\tt AffineMC} scan. We expect that this scan should be the default for simple explorations where some form of inference on the parameter distributions is desired, being able to run batches of points efficiently and scaling well to higher dimensions, and without needing to specify variances for the variables: the user need only specify their ranges. 

\subsection{Contour finding}

One common task is finding an exclusion contour. In low dimensions (typically 2 ...) this is typically performed by scanning a grid of points and interpolating. However, this can be both inefficient and inaccurate; having accurate contours would require very fine grids.

\subsubsection{Contour2D}

A simple algorithm for finding contours in two dimensions is implemented as a {\tt Contour2D} scan. It works by first creating a very coarse grid, finding the contour, and then sampling points that lie close to this contour (but not too close to each other). It performs a specified number of iterations. The contour is specified by a function, or a threshold likelihood value:
\begin{minted}
[
frame=single,
framesep=2mm,
baselinestretch=1.2,
bgcolor=White,
fontsize=\footnotesize,
mathescape=true,
linenos=false,
]
{json}
{
    "Setup": {
        "Type": "Contour2D",
        "Threshold": float=0.0,
        "Function": "<function of variables and observables>"
    }
}
\end{minted}
If the {\tt "Function"} setting is absent, it will instead use a threshold likelihood value with default $0$.

\subsubsection{ContourGP}

An interesting algorithm was proposed in \cite{lukas_heinrich_2018_1634428} to find contours using Gaussian Processes (GPs). Since GPs generalise very well from little data, they are excellent interpolators; but even better they give a prediction for the \emph{uncertainty} of each point. Exploiting this, the {\tt ContourGP} algorithm aims to choose not the points that lie closest to the exclusion contour, but the ones that give the highest \emph{entropy gain}, in other words the point that maximises the information gained about the position of the exclusion contour. The code associated with \cite{lukas_heinrich_2018_1634428}\footnote{see \url{https://github.com/diana-hep/excursion}}  does not work with recent versions of {\tt python}; the algorithm {\tt ContourGP} in \BSMArt is therefore a fresh implementation with some variations. Similar to the {\tt Contour2D} scan, it can take either a threshold likeihood value $c$ or a function $f$ of the variables and observables where the contour lies at the solution to $f=c$.

\subsection{{\sc MLScanner} and {\sc DLScanner} algorithms}

In \cite{Hammad:2022wpq}, a suite of algorithms was presented called {\sc MLScanner} along with code that can use them to run {\sc SPheno}, {\sc HiggsBounds} and {\sc HiggsSignals}. Another algorithm by the same authors was presented in \cite{Hammad:2024tzz} called {\sc DLScanner}. We ported these algorithms to \BSMArt, making modifications to improve the generality, allowing them to make use of the likelihood computations rather than focussing only on a single observable. 

There are 7 different scans under the {\sc MLScanner}. They are:
\begin{itemize}
    \item {Deep Neural Network Classifier}
    \item {Deep Neural Network Regressor}
    \item {Gradient Boosting Regressor}
    \item {Random Forest Classifier}
    \item {Random Forest Regressor}
    \item {Support Vector Regressor - Polynomial}
    \item {Support Vector Regressor - Radial Basis Function}
\end{itemize}

\subsubsection*{Gradient Boosting Regressor}

The \textsc{GBR} scan uses a Gradient Boosting Regressor from the \texttt{scikit-learn} \cite{sklearn} library to iteratively identify points with NLL (negative log likelihood) values below the user-defined threshold in the parameter space. 

The model is trained to regress the NLL of evaluated points from their 
parameter values and is used to score new candidates at each iteration. 
Unlike \textsc{MLS} (see section \ref{sec:mls}) the model is refitted on the growing dataset at 
each iteration rather than reinitialised, accumulating knowledge across 
the full run.

The scan is initiated by randomly generating \texttt{Bootstrap\_Points}  points and evaluating them. If an extra dataset is provided via \texttt{InitCSV} setting, it is merged with surviving bootstrap points. The GBR model is trained on the combined dataset to record it as a baseline.

At each iteration, a pool of \texttt{Candidate\_Points} is generated uniformly across the parameter space. The GBR model finds their NLL values and points with below \texttt{Threshold\_Value} NLL values are labeled for future evaluation. Up to
\begin{equation}
    N_\text{ML} = \lfloor \texttt{Points\_Per\_Iteration}
    \times (1 - \texttt{Random\_Fraction}) \rfloor
\end{equation}
of these are taken for evaluation, with the remainder filled with
randomly drawn points. \texttt{Random\_Fraction} acts as a minimum
exploration fraction --- if fewer promising candidates than $N_\text{ML}$
are found, the random fraction of the batch increases accordingly.
The GBR model is refitted on all valid accumulated points at each
iteration. The loop continues until \texttt{Points} new good points
have been found. The settings are listed in Table~\ref{tab:gbrsettings}
with an example configuration in Listing~\ref{lst:gbrinput}.

\begin{table}[h]
\centering
\begin{tabular}{lll}
\hline\hline
\textbf{Setting} & \textbf{Type} & \textbf{Description} \\
\hline
\texttt{Bootstrap\_Points}      & int   & Number of initial random points to evaluate \\
\texttt{Candidate\_Points}      & int   & Candidate pool size scored per iteration \\
\texttt{Points\_Per\_Iteration} & int   & Number of points evaluated per iteration \\
\texttt{Random\_Fraction}       & float & Minimum fraction of random points per batch \\
\texttt{Threshold\_Value}       & float & NLL threshold defining a good point \\
\texttt{Points}                 & int   & Number of good points to collect beyond bootstrap \\
\texttt{Estimators}             & int   & Number of boosting stages (trees) \\
\texttt{Max\_Depth}             & int   & Maximum depth of each tree \\
\texttt{LearningRate}           & float & Shrinkage factor applied to each tree \\
\texttt{Subsample}              & float & Fraction of training samples used per tree \\
\texttt{Min\_Samples\_Split}    & int   & Minimum samples required to split a node \\
\texttt{Min\_Samples\_Leaf}     & int   & Minimum samples required at a leaf node \\
\texttt{InitCSV}                & path  & Optional CSV file to seed the scan \\
\hline\hline
\end{tabular}
\caption{Settings for the \textsc{GBR} scan.}
\label{tab:gbrsettings}
\end{table}

\begin{minipage}{\linewidth}\centering
\begin{minted}
[
frame=single,
framesep=2mm,
baselinestretch=1.2,
bgcolor=LightGray,
fontsize=\footnotesize,
linenos,
]
{json}
"Networks": {
    "Bootstrap_Points":      100,
    "Candidate_Points":      5000,
    "Points_Per_Iteration":  300,
    "Random_Fraction":       0.2,
    "Threshold_Value":       1.0,
    "Estimators":            100,
    "Max_Depth":             30,
    "LearningRate":          0.01,
    "Subsample":             1.0,
    "Min_Samples_Split":     2,
    "Min_Samples_Leaf":      1
},
"Setup": {
    "RunName":       "GBR_Example",
    "Type":          "MLS_GBR",
    "Cores":         4,
    "Output File":   "output",
    "Spectrum File": "SPheno.spc.MSSM",
    "Points":        1000
}
\end{minted}
\listingcaption{Example \textsc{GBR} configuration.}\label{lst:gbrinput} 
\end{minipage}

\subsubsection*{Support Vector Regressor}

The \textsc{SVR} scans use a Support Vector Regressor from the \texttt{scikit-learn} library~\cite{sklearn} to iteratively identify points with NLL values below a user-defined threshold in the parameter space. They are based on the \textsc{MLScanner} implementationof~\cite{Hammad:2022wpq}. \textsc{SVR} scans require a kernel function to map the relationship between the input parameters and the predicted NLL. Two kernel functions are available: Radial Basis Function (\texttt{MLS\_SVR\_RBF}) and Polynomial (\texttt{MLS\_SVR\_Poly}). These two scans follow the same principles of active learning, with the only difference of being the choice of kernel functions. 

Like in other {\sc MLScanner} scans, it is initiated by evaluating the randomly generated \texttt{Bootstrap\_Points} points. The additional dataset provided via \texttt{InitCSV} is merged with the surviving bootstrap points and the SVR model is  trained on the combined dataset. The number of good points found at this stage, $N_\text{init}$ is recorded as the baseline.

At each iteration, a pool of \texttt{Candidate\_Points} is generated uniformly across the parameter space. The SVR model predicts their NLL values and candidates with predicted NLL below \texttt{Threshold\_Value} are selected for evaluation. Up to
\begin{equation}
    N_\text{ML} = \lfloor \texttt{Points\_Per\_Iteration}
    \times (1 - \texttt{Random\_Fraction}) \rfloor
\end{equation}
of these are taken for evaluation, with the remainder filled with randomly drawn points. \texttt{Random\_Fraction} acts as a minimum exploration fraction --- if fewer promising candidates than $N_\text{ML}$ are found, the random fraction of the batch increases accordingly. The SVR model is then refitted on all valid accumulated points. The loop continues until \texttt{Points} new good points have been found beyond $N_\text{init}$.

\paragraph{Radial Basis Function (\texttt{MLS\_SVR\_RBF}).}
The RBF kernel measures similarity between points using a Gaussian function of their Euclidean distance, controlled by \texttt{Gamma}, which essentially dictates whether the exploration is localised or more global, depending on larger and smaller values respectively. It is a general-purpose kernel suitable for smooth NLL landscapes without sharp boundaries.

\paragraph{Polynomial (\texttt{MLS\_SVR\_Poly}).}
The polynomial kernel maps the parameter space into a higher-dimensional
feature space via a polynomial of degree \texttt{Degree}. It can
capture more structured relationships than the RBF kernel but requires
careful tuning of \texttt{Degree} to avoid underfitting or overfitting.

The complete list of settings with examples is given in Table~\ref{tab:svr_settings} and in Listing~\ref{lst:svr_input}.
\begin{table}[h]
\centering
\begin{tabular}{llll}
\hline\hline
\textbf{Setting} & \textbf{Type} & \textbf{Applies to} & \textbf{Description} \\
\hline
\texttt{Bootstrap\_Points}      & int   & Both  & Number of initial random points to evaluate \\
\texttt{Candidate\_Points}      & int   & Both  & Candidate pool size scored per iteration \\
\texttt{Points\_Per\_Iteration} & int   & Both  & Number of points evaluated per iteration \\
\texttt{Random\_Fraction}       & float & Both  & Minimum fraction of random points per batch \\
\texttt{Threshold\_Value}       & float & Both  & NLL threshold defining a good point \\
\texttt{Points}                 & int   & Both  & Number of good points to collect beyond bootstrap \\
\texttt{C}                      & float & Both  & Regularisation parameter \\
\texttt{Gamma}                  & float & Both  & Kernel length scale parameter \\
\texttt{Epsilon}                & float & Both  & Width of the epsilon-insensitive loss tube \\
\texttt{Degree}                 & int   & Poly only & Degree of the polynomial kernel \\
\texttt{InitCSV}                & path  & Both  & Optional CSV file to seed the scan \\
\hline\hline
\end{tabular}
\caption{Settings for the \textsc{SVR} scans.}
\label{tab:svr_settings}
\end{table}

\begin{minipage}{\linewidth}\centering
\begin{minted}
[
frame=single,
framesep=2mm,
baselinestretch=1.2,
bgcolor=LightGray,
fontsize=\footnotesize,
linenos,
]
{json}
"Networks": {
    "Bootstrap_Points":      100,
    "Candidate_Points":      5000,
    "Points_Per_Iteration":  300,
    "Random_Fraction":       0.2,
    "Threshold_Value":       1.0,
    "C":                     100,
    "Gamma":                 0.1,
    "Epsilon":               0.1,
    "Degree":                3
},
"Setup": {
    "RunName":       "SVR_Example",
    "Type":          "MLS_SVR_RBF",
    "Cores":         4,
    "Output File":   "output",
    "Spectrum File": "SPheno.spc.MSSM",
    "Points":        1000
}
\end{minted}
\listingcaption{Example \textsc{SVR} configuration. For the polynomial kernel,
set \texttt{Type} to \texttt{MLS\_SVR\_Poly} and include the
\texttt{Degree} setting.}
\label{lst:svr_input}
\end{minipage}

\subsubsection*{Random Forest Scans}

The Random Forest scans are other methods using the \texttt{scikit-learn} library~\cite{sklearn}. They follow the same logic as other \textsc{MLScanner}\cite{Hammad:2022wpq} based scans. Two variants are available: a Regressor (\texttt{MLS\_RFR}) and a Classifier (\texttt{MLS\_RFC}). Both share identical active learning logic involving \texttt{Bootstrap\_Points}, \texttt{InitCSV}, \texttt{Points}, \texttt{Candidate\_Points} and \texttt{Random\_Fraction} settings, as described for the \textsc{GBR} and \textsc{SVR} scans.

The RFR model is trained to regress the NLL of evaluated points from the scan variables. Candidates are selected if their predicted NLL falls below \texttt{Threshold\_Value}, following the same hard threshold selection as \textsc{GBR} and \textsc{SVR}.

The RFC model reframes the problem as binary classification, labelling points as good or bad using \texttt{Threshold\_Value} as the decision boundary. Before training, NLL values are converted to according labels, by using the \texttt{Threshold\_Value} as the ground. The model gets trained to predict the label of the candidate points and choose the top candidates for physics evaluation.

The complete list of the Random Forest scan settings with their examples are given in Table \ref{tab:rfsettings} and in Lists \ref{lst:rfcinput}, \ref{lst:rfrinput}.

\begin{table}[h]
\centering
\begin{tabular}{llll}
\hline\hline
\textbf{Setting} & \textbf{Type} & \textbf{Default} & \textbf{Description} \\
\hline
\texttt{Bootstrap\_Points}      & int   & 100       & Number of initial random points to evaluate \\
\texttt{Candidate\_Points}      & int   & 500       & Candidate pool size scored per iteration \\
\texttt{Points\_Per\_Iteration} & int   & 300       & Number of points evaluated per iteration \\
\texttt{Random\_Fraction}       & float & 0.2       & Minimum fraction of random points per batch \\
\texttt{Threshold\_Value}       & float & 1.0       & NLL threshold defining a good point \\
\texttt{Points}                 & int   & 10000     & Number of good points to collect beyond bootstrap \\
\texttt{Estimators}             & int   & 100 / 300 & Number of trees (RFR / RFC) \\
\texttt{Max\_Depth}             & int   & 30 / 50   & Maximum depth of each tree (RFR / RFC) \\
\texttt{Min\_Samples\_Split}    & int   & 2         & Minimum samples required to split a node \\
\texttt{Min\_Samples\_Leaf}     & int   & 1         & Minimum samples required at a leaf node \\
\texttt{InitCSV}                & path  & ---       & Optional CSV file to seed the scan \\
\hline\hline
\end{tabular}
\caption{Settings for the Random Forest scans. Where two defaults
are listed, the first applies to \texttt{MLS\_RFR} and the second
to \texttt{MLS\_RFC}.}
\label{tab:rfsettings}
\end{table}

\begin{minipage}{\linewidth}\centering
\begin{minted}
[
frame=single,
framesep=2mm,
baselinestretch=1.2,
bgcolor=LightGray,
fontsize=\footnotesize,
linenos,
]
{json}
"Setup": {
    "RunName":       "RFR_Example",
    "Type":          "MLS_RFR",
    "Cores":         4,
    "Output File":   "output",
    "Spectrum File": "SPheno.spc.MSSM",
    "Points":        500
},
"Networks": {
    "Threshold_Value":       1,
    "Bootstrap_Points":      100,
    "Candidate_Points":      500,
    "Points_Per_Iteration":  300,
    "Random_Fraction":       0.2,
    "Estimators":            100,
    "Max_Depth":             30,
    "Min_Samples_Split":     2,
    "Min_Samples_Leaf":      1
}
\end{minted}
\listingcaption{Example \texttt{MLS\_RFR} configuration.}
\label{lst:rfrinput}
\end{minipage}

\begin{minipage}{\linewidth}\centering
\begin{minted}
[
frame=single,
framesep=2mm,
baselinestretch=1.2,
bgcolor=LightGray,
fontsize=\footnotesize,
linenos,
]
{json}
"Setup": {
    "RunName":       "RFC_Example",
    "Type":          "MLS_RFC",
    "Cores":         4,
    "Output File":   "output",
    "Spectrum File": "SPheno.spc.MSSM",
    "Points":        500
},
"Networks": {
    "Threshold_Value":       1,
    "Bootstrap_Points":      100,
    "Candidate_Points":      500,
    "Points_Per_Iteration":  300,
    "Random_Fraction":       0.2,
    "Estimators":            300,
    "Max_Depth":             50,
    "Min_Samples_Split":     2,
    "Min_Samples_Leaf":      1
}
\end{minted}
\listingcaption{Example \texttt{MLS\_RFC} configuration.}
\label{lst:rfcinput}
\end{minipage}

\subsubsection*{Deep Neural Network Scans}

The Deep Neural Network scans use \texttt{PyTorch}~\cite{pytorch} based neural networks to iteratively identify points with NLL values below a user-defined threshold. 
Two variants are available: a Regressor (\texttt{MLS\_DNNR}) and a Classifier (\texttt{MLS\_DNNC}). Both share identical active learning logic involving \texttt{Bootstrap\_Points}, \texttt{InitCSV}, \texttt{Points}, \texttt{Candidate\_Points} and \texttt{Random\_Fraction} settings, as described for other scans in this class.

Both variants share an adaptive architecture selector that chooses the network structure based on the size of the training dataset. When the number of training points exceeds 5000, a funnel architecture with SiLU activation and Dropout(0.2) is used:
\begin{equation}
    \mathbb{R}^n \to 2h \xrightarrow{\text{SiLU, Dropout}}
    h \xrightarrow{\text{SiLU, Dropout}}
    \frac{h}{2} \xrightarrow{\text{SiLU, Dropout}} \mathbb{R}^1,
\end{equation}
where $h$ is set by \texttt{Neurons} and $n$ is the number of scan variables. When fewer than 5000 training points are available, a flat architecture with ReLU activation and no dropout is used instead:
\begin{equation}
    \mathbb{R}^n \to h \xrightarrow{\text{ReLU}}
    h \xrightarrow{\text{ReLU}}
    h \xrightarrow{\text{ReLU}}
    h \xrightarrow{\text{ReLU}}
    h \xrightarrow{\text{ReLU}} \mathbb{R}^1.
\end{equation}
The flat architecture is better suited to small datasets where the more expressive funnel would overfit.

The DNNR network takes the scan variables as input and predicts the corresponding NLL using MSE loss. Candidates are selected if their predicted NLL falls below \texttt{Threshold\_Value}, following the same hard threshold selection as regressor based methods in this class.

The DNNC network is a binary classification based scan, which labels points as good or bad using \texttt{Threshold\_Value} as the decision boundary, and is trained with binary cross-entropy loss. The network uses the same confidence-ranked candidate selection as \texttt{MLS\_RFC}.

When good points are sparse in a given iteration, the DNNC scan applies the Synthetic Minority Over-sampling Technique (SMOTE)~\cite{smote} to augment the training data. Two regimes are defined based on the number of new good points $n_+$ found in the current batch:

\paragraph{Very sparse regime ($n_+ \leq$ \texttt{Kinitial}).}
The new good points are pooled with randomly selected points from the accumulated good dataset to form a SMOTE input of size $2 \times \texttt{Kinitial}$. SMOTE generates synthetic points by interpolating between nearest neighbours in this pool, which are then added to the good dataset if they pass the threshold.

\paragraph{Moderately sparse regime ($n_+>$ \texttt{Kinitial} and $n_+$ less than half the ML-selected quota).}

SMOTE is applied directly to the good and bad points found in the current batch, without drawing from the accumulated dataset. 

When neither condition is met, the batch is used directly without oversampling. In all cases, good points found through SMOTE contribute to the stopping criterion.

The complete list of settings with examples is given in Table~\ref{tab:dnnsettings} and Listings~\ref{lst:dnnrinput} and~\ref{lst:dnncinput}.

\begin{table}[h]
\centering
\begin{tabular}{lll}
\hline\hline
\textbf{Setting} & \textbf{Type} & \textbf{Description} \\
\hline
\texttt{Bootstrap\_Points}      & int   & Number of initial random points to evaluate \\
\texttt{Candidate\_Points}      & int   & Candidate pool size scored per iteration \\
\texttt{Points\_Per\_Iteration} & int   & Number of points evaluated per iteration \\
\texttt{Random\_Fraction}       & float & Minimum fraction of random points per batch \\
\texttt{Threshold\_Value}       & float & NLL threshold defining a good point \\
\texttt{Points}                 & int   & Number of good points to collect beyond bootstrap \\
\texttt{Neurons}                & int   & Width parameter $h$ for the network architecture \\
\texttt{Epochs}                 & int   & Maximum training epochs per iteration \\
\texttt{LearningRate}           & float & Learning rate for the AdamW optimiser \\
\texttt{Batch Size}             & int   & Mini-batch size for training \\
\texttt{Kinitial}               & int   & SMOTE neighbourhood size (\texttt{MLS\_DNNC} only) \\
\texttt{InitCSV}                & path  & Optional CSV file to seed the scan \\
\hline\hline
\end{tabular}
\caption{Settings for the Deep Neural Network scans. \texttt{Kinitial}
applies to \texttt{MLS\_DNNC} only.}
\label{tab:dnnsettings}
\end{table}

\begin{minipage}{\linewidth}\centering
\begin{minted}
[
frame=single,
framesep=2mm,
baselinestretch=1.2,
bgcolor=LightGray,
fontsize=\footnotesize,
linenos,
]
{json}
"Setup": {
    "RunName":       "DNNR_Example",
    "Type":          "MLS_DNNR",
    "Cores":         4,
    "Output File":   "output",
    "Spectrum File": "SPheno.spc.MSSM",
    "Points":        500
},
"Networks": {
    "Threshold_Value":       1,
    "Bootstrap_Points":      10,
    "Candidate_Points":      5,
    "Points_Per_Iteration":  5,
    "Random_Fraction":       0.2,
    "Neurons":               100,
    "Epochs":                1000,
    "LearningRate":          0.01,
    "Batch Size":            5
}
\end{minted}
\listingcaption{Example \texttt{MLS\_DNNR} configuration. Note that the
default values in the code are set for testing purposes and should
be increased for production runs.}
\label{lst:dnnrinput}
\end{minipage}

\begin{minipage}{\linewidth}\centering
\begin{minted}
[
frame=single,
framesep=2mm,
baselinestretch=1.2,
bgcolor=LightGray,
fontsize=\footnotesize,
linenos,
]
{json}
"Setup": {
    "RunName":       "DNNC_Example",
    "Type":          "MLScanner.MLS_DNNC",
    "Cores":         4,
    "Output File":   "output",
    "Spectrum File": "SPheno.spc.MSSM",
    "Points":        500
},
"Networks": {
    "Threshold_Value":       1,
    "Bootstrap_Points":      100,
    "Candidate_Points":      500,
    "Points_Per_Iteration":  300,
    "Random_Fraction":       0.2,
    "Neurons":               100,
    "Epochs":                250,
    "LearningRate":          0.01,
    "Batch Size":            500,
    "Kinitial":              5
}
\end{minted}
\listingcaption{Example \texttt{MLS\_DNNC} configuration.}
\label{lst:dnncinput}
\end{minipage}

\subsection{\textsc{MLS}}
\label{sec:mls}

The \textsc{MLS} scan is based on the scan of the same name implemented
in \textsc{xBit}~\cite{Staub:2019xhl}, which applied the approach
originally proposed in~\cite{Ren:2017mls}. It is implemented using the
\texttt{PyTorch} library. The fundamental idea is to train a neural
network on previously evaluated points and use it to propose new
candidates in promising regions of parameter space. 

The predictor network architecture is configured by the user via the
\texttt{Neurons} setting, which specifies the widths of the hidden
layers $[h_1, h_2, \ldots, h_k]$. Between each hidden layer, ReLU
activation is applied followed by dropout with a default rate of 0.1.
The output dimension $d_\text{out}$ depends on the training mode
controlled by the \texttt{TrainLH} setting: when \texttt{TrainLH} is
\texttt{True} the network trains on the likelihood and $d_\text{out}
= 1$, whereas when \texttt{TrainLH} is \texttt{False} the network
trains on the observables directly and $d_\text{out} = n_\text{obs}$,
the number of active observables. At the start of every iteration, the
network is reinitialised with fresh random weights and trained from
scratch on the full set of valid points accumulated so far.

When \texttt{Classifier} is set to \texttt{True}, a second network is
trained to distinguish physically valid from invalid parameter points.
During candidate proposal, only points predicted as valid by the
classifier are passed forward for likelihood scoring.

At each iteration, a large pool of candidates is drawn uniformly from
the parameter space. If the classifier is enabled, candidates predicted
as invalid are discarded first. The remaining candidates are then scored
by the predictor network. If the density penalty is enabled, it is
applied to the predicted likelihoods at this stage (see below).

From the scored candidates, the maximum predicted likelihood
$\ell_\text{best}$ is identified. Points are then selected for
evaluation if their predicted likelihood $\ell$ satisfies
\begin{equation}
    0.1 \cdot \ell_\text{best} \;<\; \ell \;<\; 2,
\end{equation}
where the upper bound of 2 accommodates cases where the predictor
overestimates the likelihood of good points. This selection is repeated
for up to 10 passes until 90\% of \texttt{Points} are collected, with
the candidate pool regenerated at each pass if needed. The remaining
10\% of the batch is always filled with uniformly random points to
maintain exploration. If no candidates survive scoring, the full batch
falls back to random points.

\texttt{DensityPenalty} is a boolean switch which is active by default.
When enabled, it penalises candidates that lie close to already-evaluated
points, encouraging broader exploration of the parameter space. It can
be disabled by setting \texttt{DensityPenalty} to \texttt{False}.

The full list of settings for the \textsc{MLS} scan is given in
Table~\ref{tab:mls_settings}, with an example configuration in
Listing~\ref{lst:mls_input}.

\begin{table}[h]
\centering
\begin{tabular}{lll}
\hline\hline
\textbf{Setting} & \textbf{Type} & \textbf{Description} \\
\hline
\texttt{LR}             & float      & Learning rate for the Adam optimiser \\
\texttt{Neurons}        & list[int]  & Hidden layer widths, e.g.\ \texttt{[25,25,25]} \\
\texttt{Points}         & int        & Number of points evaluated per iteration \\
\texttt{Iterations}     & int        & Number of active learning cycles \\
\texttt{Epochs}         & int        & Maximum training epochs per iteration \\
\texttt{TrainLH}        & bool       & Learn likelihood (\texttt{True}) or observables (\texttt{False}) \\
\texttt{LogLike}        & bool       & Use log-likelihood for training and scoring \\
\texttt{Classifier}     & bool       & Enable the optional validity classifier \\
\texttt{DensityPenalty} & bool       & Apply distance penalty during candidate selection \\
\hline\hline
\end{tabular}
\caption{Settings for the \textsc{MLS} scan.}
\label{tab:mls_settings}
\end{table}

\begin{minipage}{\linewidth}\centering
\begin{minted}
[
frame=single,
framesep=2mm,
baselinestretch=1.2,
bgcolor=LightGray,
fontsize=\footnotesize,
linenos,
]
{json}
"Setup": {
    "RunName":        "MLS_MSSM",
    "Type":           "MLS",
    "Cores":          4,
    "Output File":    "MSSM_Output",
    "Spectrum File":  "SPheno.spc.MSSM",
    "LR":             0.001,Example
    "Neurons":        [25, 25, 25],
    "Points":         100,
    "Iterations":     100,
    "Epochs":         5000,
    "TrainLH":        "False",
    "Classifier":     "True",
    "LogLike":        "True",
    "DensityPenalty": "False"
}
\end{minted}
\listingcaption{Example \textsc{MLS} configuration.\label{lst:mls_input}}
\end{minipage}

\section{CMA-ES scans}
\label{SEC:CMAES}

We implement in BSMArt the AI black-box optimisation strategy introduced in ref.~\cite{deSouza:2022uhk}, which was later extended with a novelty reward mechanism in ref.~\cite{Romao:2024gjx} and with a hierarchical loss in ref.~\cite{deSouza:2025bpl}. This approach has since been successfully applied to a large array of BSM problems~\cite{Basiouris:2024qfe,Boto:2025mmn,Boto:2025ovp,deSouza:2025uxb,Boto:2026gzj,deSouza:2026sta}. The methodology reframes parameter space scans as an optimisation problem. Namely, it relies on minimising a constraint function, $C$, defined as
\begin{equation}
    C(\mathcal{O}) = \max{(0, - \mathcal{O}+\mathcal{O}_{\textrm{LB}}, \mathcal{O}-\mathcal{O}_{\textrm{UB}})},
    \label{eq:constraint_func}
\end{equation}
where $\mathcal{O}$ is the predicted value for the observable considered\footnote{Or other constrained quantity, e.g. self-consistency or virtually any quantifiable user-defined criterion.} and $\mathcal{O}_{\textrm{LB}}$ and $\mathcal{O}_{\textrm{UB}}$ are the experimental (or theoretical, when applied) lower and upper bounds, respectively, for the observable in consideration.\footnote{These need not to be finite, e.g. they can be used to set one-side unbounded constraints.} 
Notably, $C(\mathcal{O})$ returns 0 if and only if $\mathcal{O}$ lies within the specified interval; otherwise, it returns a positive number quantifying \textit{how far} the value of the observable is from the specified bounds. This ensures that even points that fail to satisfy a given constraint can still provide useful information to be leveraged by the optimisation algorithm. For $N_C$ such constraints, a single \emph{loss function}\footnote{See~\cite{deSouza:2025uxb} for an alternative approach using multi-objective search algorithms.} is computed by summing over all individual constraints\footnote{In practice, each contribution is first scaled via $C\to \log(1+C)$ and then they are min-maxed generation-wise before aggregating through summation. These steps guarantee the existence of a numerical infinity for \emph{deadly} penalties and prevent one constraint from dominating the loss function.}
\begin{equation}
    L = \sum_{i=1}^{N_C} C(\mathcal{O}_i) \ .
\end{equation}

In ref.~\cite{Romao:2024gjx} the loss function was modified to introduce novelty detection rewards. To achieve this, we first increase the loss function by an offset of $1$ outside the validity region
\begin{equation}
	\tilde L =\begin{cases}
		1 +L & \text{if } L > 0 \\
		0            & \text{if } L = 0
	\end{cases} \ ,
\end{equation}
then, we add new terms for $N_p$ penalties
\begin{equation}
	L_T = \tilde L + \frac{1}{N_p}\sum_{i=1}^{N_p} p_i \ ,
\end{equation}
where the penalties need to be normalised so that $0\leq p_i\leq1$. With these two changes, we guarantee that the region of validy is still defined by the vanishing of the constraint functions and prevent a competition between constraints and penalties during the optimisation run. Since $ \sum_i C(\mathcal{O}_i) = 0 \Rightarrow 0 \leq L_T \leq 1$, the novelty reward extension allows for the scanner to keep exploring the valid parameter (or observable) space after convergence. This will force the algorithm to find diverse minima, i.e. to chart the subspace of validity and has shown to efficiently find solutions with previously unrealised phenomenology~\cite{Romao:2024gjx, deSouza:2025bpl, deSouza:2025uxb}. In this version of BSMArt, we include the \texttt{"HBOS"} penalty used in these studies, which is density-based and forces CMA-ES to explore away from regions of high density, i.e. of regions that it has already explored.

For this version of BSMArt, the optimisation algorithm of choice is CMA-ES~\cite{CMAES-original, nomura2026cmaessimplepracticalpython} as it has shown to converge often and quickly to viable regions of highly constrained multidimensional parameter spaces~\cite{deSouza:2022uhk, Romao:2024gjx}. More details about the CMA-ES in HEP are described in refs.~\cite{deSouza:2022uhk, Romao:2024gjx, deSouza:2025uxb, deSouza:2025bpl}. For convenience, two separate CMA-ES scans are provided that cover different use cases; \verb|"CMAES"| and \verb|"CMAES_ND"|, where \verb|"CMAES"| contains the basic implementation and \verb|"CMAES_ND"| offers the novelty reward enhancement:
\begin{itemize}
    \item \verb|"CMAES"|: The basic implementation provides a powerful minimisation algorithm applied to the negative log likelihood.  
    As was shown in~\cite{Basiouris:2024qfe}, this will efficiently find the modes of maximal posterior probability. Indeed, CMA-ES has been shown to be related to \emph{natural gradient descent}, which is the statistically optimal way of finding Maximum A Posteriori modes~\cite{akimoto2012theoretical}. 
    \item \verb|"CMAES_ND"|: The novelty detection variation, which also includes hierarchical loss, allows for continual exploration after convergence. This makes it ideal to systematically explore hypersurfaces of validity or to chart the diversity of possible phenomenological signatures. For better exploration , the use of \verb|"CMAES_ND"| is recommended.
\end{itemize}

When using 
\verb|"CMAES_ND"| 
as the scan of choice it is necessary to use the \verb|"CFUNCTION"| scaling for all the observables to be optimised by the scan. Additionally, it is recommended to use the \verb|"VALID"| option for some of the observables which ought to be used as validity conditions. This ensures the HEP tool chain is interrupted every time one of the observables of choice is outside of the specified range. This effectively saves computation time and speeds up convergence.

In this work, alongside the density-based novelty detection reward employed in previous studies using \texttt{"HBOS"}, we introduce a new novelty detection option to be used in the CMA-ES scans (\verb|"CMAES_ND"|) that aims at optimising a specific quantity. This uses a sigmoid function $\sigma(\mathcal{O}_i) = \frac{1}{1+e^{\pm\mathcal{O}_i}}$ to maximise (or minimise) observables of choice ($\mathcal{O}_i$), for example; cross-sections, branching ratios, $\chi^{2}$ test, etc. This function can be accessed by setting \verb|"CMAESNoveltyDetection"| to \texttt{"Optimiser"} in the \verb|"Setup"| section of the \verb|json| file.

The new settings associated with the \verb|"CMAES_ND"| scan are:
\begin{itemize}
    \item  \verb|"CMAESPopulationSize"|: Number of points to run per batch in parallel (optional, default: $4 + \floor{3 \log{n_{\text{variables}}}}$).
    \item  \verb|"CMAESMean"|: Initial mean value for CMA-ES scan (default: random vector in $[0.0, 1.0]^{n_{\text{variables}}}$).
    \item  \verb|"CMAESSigma"|: Initial step size for CMA-ES scan (optional, default: $1/\sqrt{n_{\text{variables}}}$).
    \item  \verb|"CMAESMaxGenerations"|: Number of total batches to run (default: 100).
    \item  \verb|"CMAES MaxLoss"|: Override the default maximum loss value for points that fail to run a given HEP tool (optional).
    \item  \verb|"CMAESPathSeeds"|: Path to file (.csv or .parquet) containing seeds for initial mean (optional).
    \item  \verb|"CMAESNoveltyDetection"|: Specify the novelty detection algorithm to use. Options: \texttt{"HBOS"} or \texttt{"Optimiser"}.
    \item \verb|"CMAESNoveltyDetectionParams"|: Dictionary specifying the hyperparameters for novelty detection. Options: \verb|"n_bins"| (number of bins for HBOS) and \verb|"direction"| (direction for Optimiser. Can be \verb|"max"|, to maximise an observable, or \verb|"min"|, to minimise it.)
    \item  \verb|"CMAESFocus"|: Dictionary with \verb|"Variables"| and \verb|"Observables"| lists for novelty detection focus (optional).
    \item  \verb|"CMAESHierarchy"|: List of hierarchical observables to be prioritised during optimisation (optional).
    \item  \verb|"CMAESNoGoodPointsPatience"|: Maximum number of generations without good points before stopping (default: 5000).
    \item  \verb|"CMAESGoodPointsEarlyStop"|: Target number of good points for early stopping (default: 5000).
    \item  \verb|"CMAESBestLossPatience"|: Patience for best loss counter. Maximum number of generations without improvement to the loss function (default: 200).
    \item  \verb|"CMAESBestNValidConstraintsPatience"|: Patience for best number of valid constraints counter. Maximum number of generations without improvement in the number of constraints satisfied (default: 2000).
\end{itemize}

An example of \verb|"CMAES_ND"| scan is given in the \verb|BSMArt_pMSSM_CMAES.json| file 
\begin{minted}
[
frame=single,
framesep=2mm,
baselinestretch=1.2,
bgcolor=White,
fontsize=\footnotesize,
mathescape=true,
linenos=false,
]
{json}
{
    "Codes": {
        "SPheno": {
            ...
            "Observables": {
                "mh": {
                    "SLHA": ["MASS", [25]],
                    "SCALING": "CFUNCTION",
                    "MEAN": 125.09,
                    "VARIANCE": 3,
                    "VALID": [122.09,128.09]
                },
                "BR_Chi_2_Fe_1": {
                    "SLHA": ["BR", [1000023, [-11, 11, 1000022]]],
                    "SCALING": "OFF"
                },
                "BR_Chi_2_Fe_2": {
                    "SLHA": ["BR", [1000023, [-13, 13, 1000022]]],
                    "SCALING": "OFF"
                },
                "sum_BR_Chi_2": {
                    "FUNCTION": "BR_Chi_2_Fe_1 + BR_Chi_2_Fe_2",
                    "SCALING": "CFUNCTION",
                    "LOWER_BOUND": 0.07,
                    "VALID": [0.07,"inf"]
                }
            },
             "Run": "True"
        },
        ...
    "Setup": {
        "RunName": "pMSSM-optimiser-BR",
        "Type": "CMAES_ND",
        ...
        "LogLike": "True",
        "CMAESPopulationSize": null,
        "CMAESMean": null,
        "CMAESMaxGenerations": 10000,
        "CMAES MaxLoss": null,
        "CMAESNoveltyDetection": "Optimiser",
        "CMAESNoveltyDetectionParams": {"direction": "max"},
        "CMAESFocus": {
            "Variables": [],
            "Observables": ["sum_BR_Chi_2"]
        },
        "CMAESHierarchy": [],
        "CMAESNoGoodPointsPatience": 1000,
        "CMAESGoodPointsEarlyStop": 1000,
        "CMAESBestLossPatience": 1000,
        "CMAESBestNValidConstraintsPatience": 1000
    },
    ...
}
\end{minted}

\section{Examples}
\label{SEC:PHYSICS}

\subsection{pMSSM scan with CMA-ES ND}

As discussed in the introduction, CMS and ATLAS both observed excesses in soft lepton and monojet searches \cite{ATLAS:2019lng,ATLAS:2021moa,CMS:2021edw,ATLAS:2021kxv,CMS:2021far}. There were subsequently CMS and ATLAS interpretations of the soft lepton excesses in the pMSSM~\cite{CMS-PAS-SUS-24-004,ATLAS:2024qmx}. 
Motivated by these, and the \smodels analysis \cite{Constantin:2025bqp}, 
we perform three simple example \verb|"CMAES_ND"| scans in the pMSSM parameter space. 
The evaluation pipeline include \SPheno to calculate the mass spectrum and decays branching ratios, {\sc MicrOMEGAS} to calculate dark matter observables and \smodels to verify exclusion limits from LHC analyses. 
The \SPheno and {\sc MicrOMEGAS} codes are generated using \SARAH. More information about parameter ranges and constraints on observables is available in appendix~\ref{app:scan}.

The three examples are aimed at covering a variety of use cases for the \verb|"CMAES_ND"| scan which should be part of any global scan strategy. The first example is a simple general pMSSM scan with the aim of producing an exploratory cover of different phenomenological results. Then, a collection of interesting points generated from the first scan is selected to be used as seeds for subsequent scans. The second scan uses \texttt{"HBOS"} for novelty detection to produce a comprehensive chart of possible phenomenology, whereas the third one uses \texttt{"Optimiser"} to produce viable points with a large value of a specific quantity.

\subsubsection{General exploratory scan}

We first perform a general exploratory scan consisting of 1000 independent runs of \verb|"CMAES_ND"| working in parallel. 
A single run by itself often converges quickly to a valid region of the parameter space due to the exploitative and local nature of CMA-ES, albeit covering a limited region of the parameter space. 
Therefore, a collection of independent parallel runs is often necessary to ensure a good initial map of interesting regions in the parameter space. 
This strategy greatly increases the coverage and efficiency of any exploratory analysis, and is strongly advised. 
For this initial scan, we terminated each run as soon as they converged to a region of validity, i.e.\ when a point is found for which all constraints in~\cref{app:pmssm_constraints} are met.

\begin{figure}[H]
	\centering
	\includegraphics[width = 0.45\textwidth]{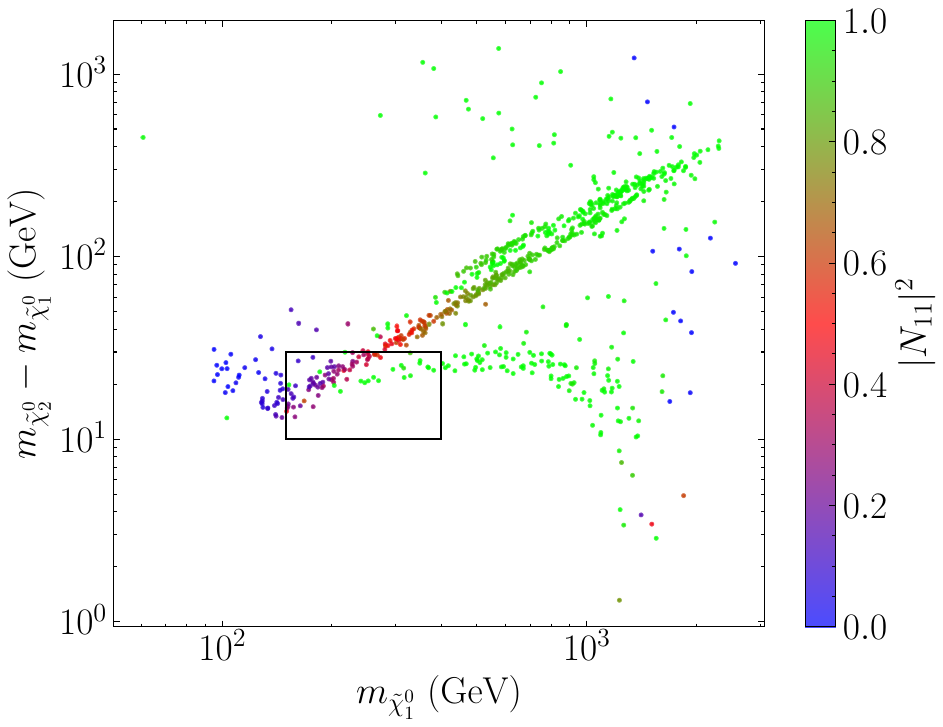}
    \includegraphics[width = 0.45\textwidth]{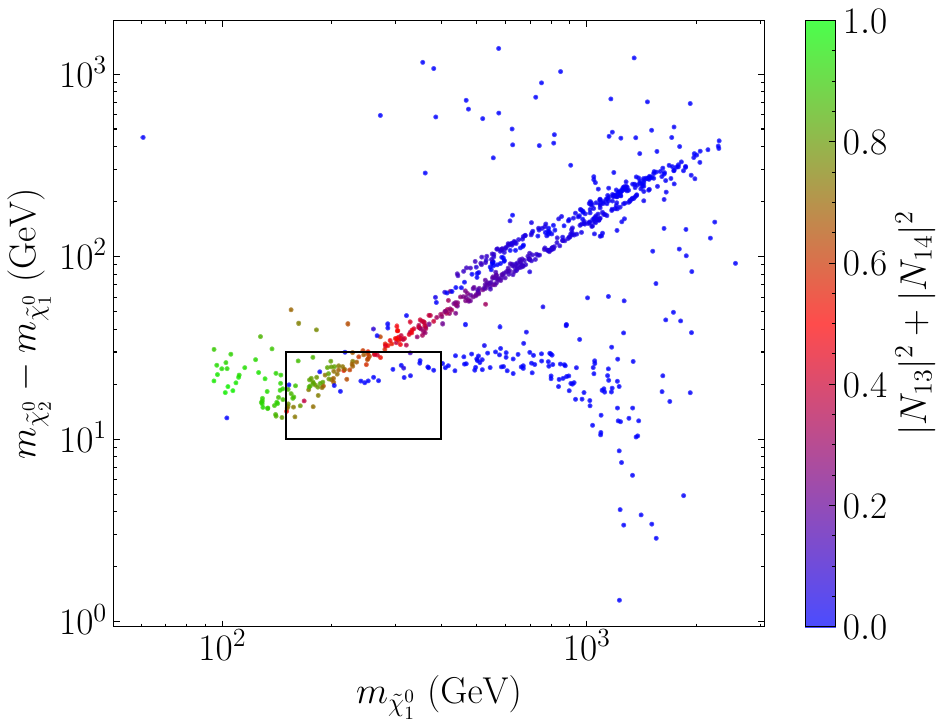}
	\caption{LSP mass and mass difference between LSP and the lightest neutralino state. Points obtained from general CMAES-ND scan and are color coded by the bino content (left) and higgsino content (right).}
	\label{fig:general_scan_m_Chi_1_Delta_m_Chi_1_Chi_2}
\end{figure}

\begin{figure}[H]
	\centering
	\includegraphics[width = 0.45\textwidth]{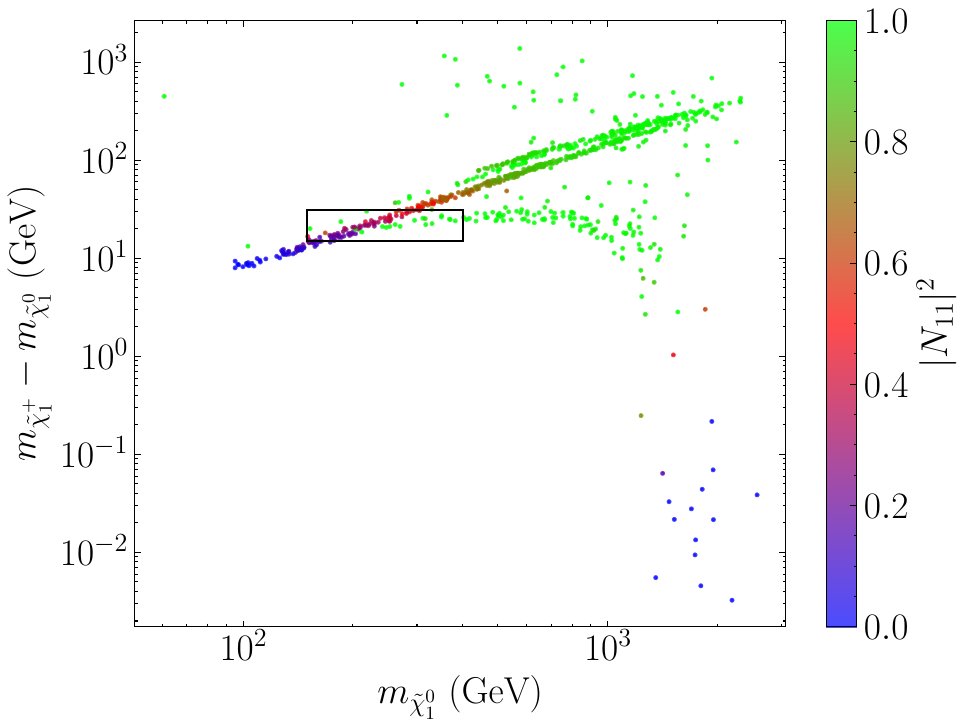}
    \includegraphics[width = 0.45\textwidth]{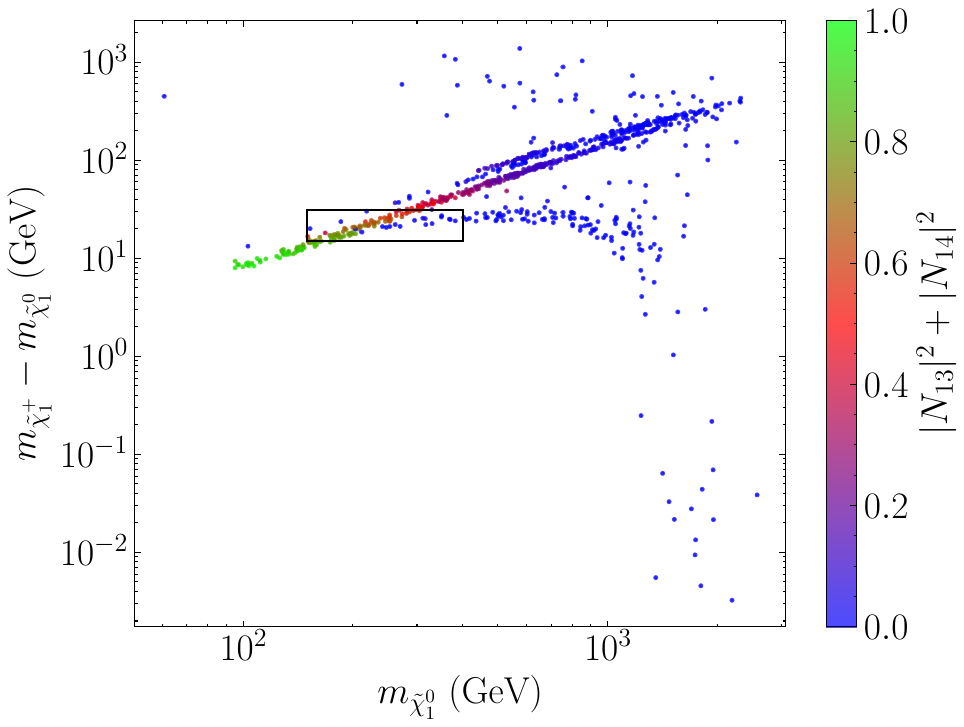}
	\caption{LSP mass and mass difference between LSP and the lightest chargino state. Points obtained from general CMAES-ND scan and are color coded by the bino content (left) and higgsino content (right). 
    }
	\label{general_scan_m_Chi_1_Delta_m_Chi_1_Cha_1}
\end{figure}

The results of this general exploratory scan are presented in~\cref{fig:general_scan_m_Chi_1_Delta_m_Chi_1_Chi_2,general_scan_m_Chi_1_Delta_m_Chi_1_Cha_1}, where we present the scatters in the planes  $(m_{\tilde{\chi}^{0}_1}, m_{\tilde{\chi}^{0}_2} - m_{\tilde{\chi}^{0}_1})$ and $(m_{\tilde{\chi}^{0}_2}, m_{\tilde{\chi}^{0}_2} - m_{\tilde{\chi}^{0}_1})$. We identify a region of interest with $150 \text{ GeV}< m_{\tilde{\chi}^{0}_1} < 400 \text{ GeV}$ and $10 \text{ GeV}< m_{\tilde{\chi}^{0}_2} - m_{\tilde{\chi}^{0}_1} < 30 \text{ GeV}$. These are motivated by by the experimental analyses \cite{ATLAS:2021moa,CMS:2021edw,ATLAS:2021kxv}
where an excess in the lepton channel has been observed.
We draw seeds from these regions for the focused scans of the next sections.

\subsubsection{Seeded HBOS ND scan}

\begin{figure}[H]
	\centering
	\includegraphics[width = 0.45\textwidth]{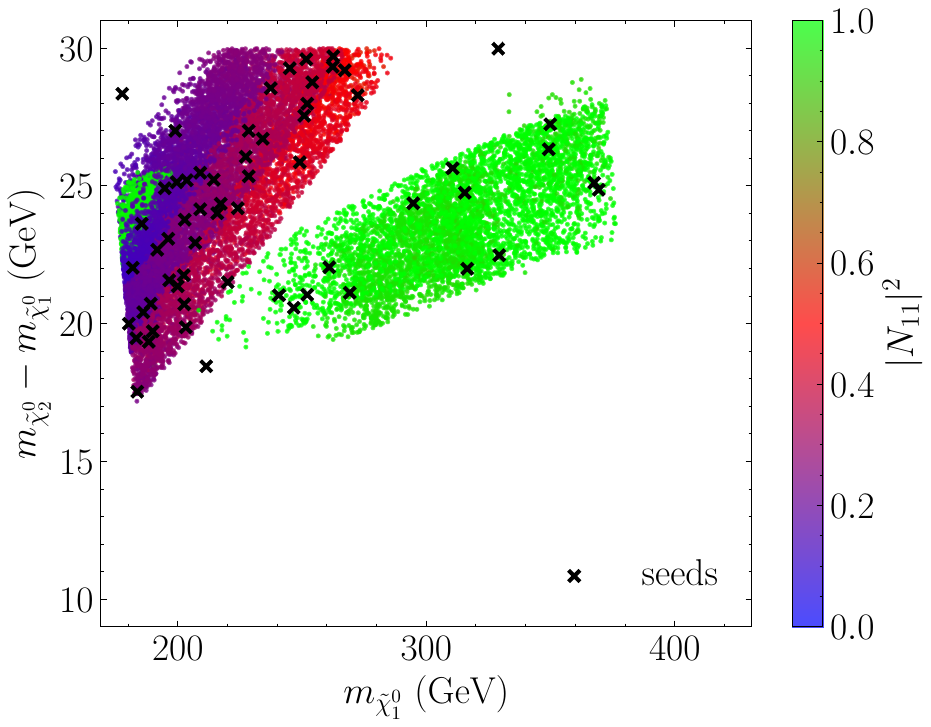}
    \includegraphics[width = 0.45\textwidth]{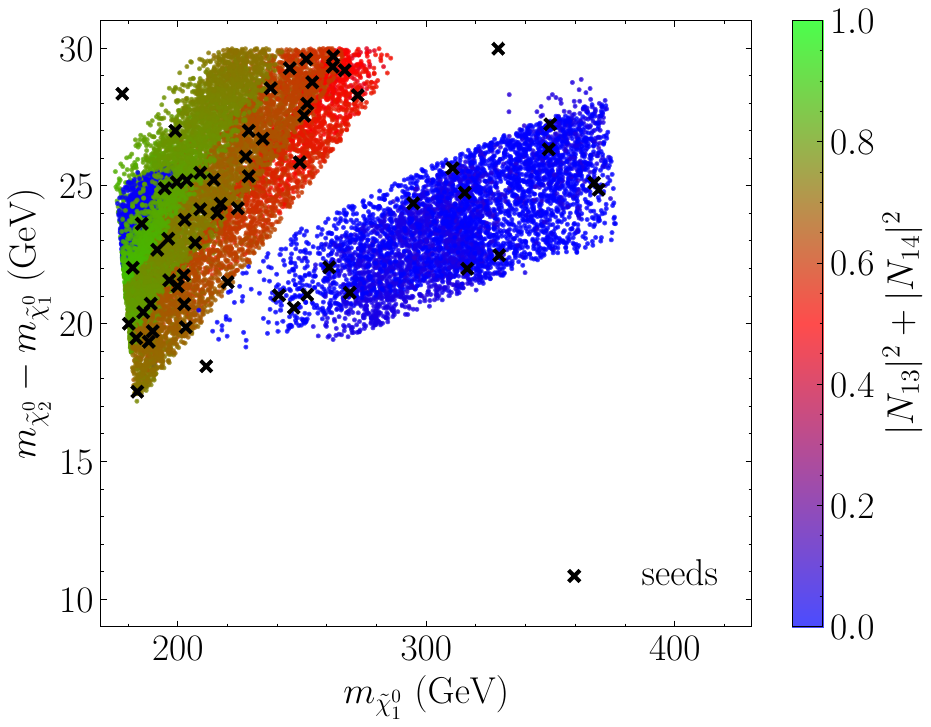}
	\caption{LSP mass and mass difference between LSP and the lightest neutralino state. Points obtained from seeded CMAES-ND scan and are color coded by the bino content (left) and higgsino content (right).
    }
	\label{fig:seeded_scan_m_Chi_1_Delta_m_Chi_1_Chi_2}
\end{figure}

\begin{figure}[H]
	\centering
	\includegraphics[width = 0.45\textwidth]{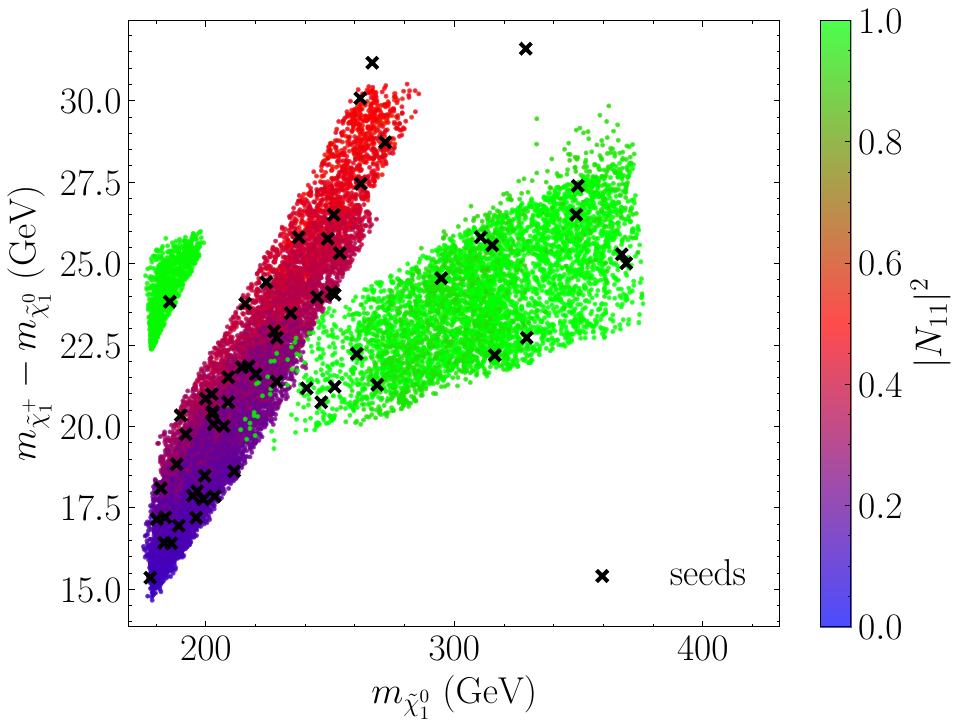}
    \includegraphics[width = 0.45\textwidth]{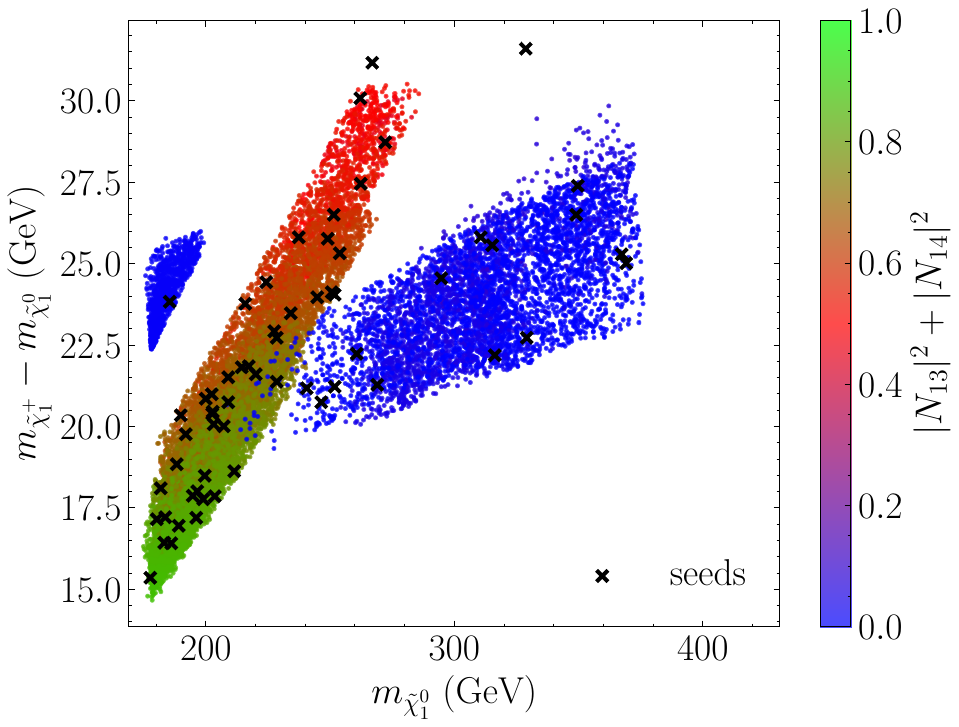}
	\caption{LSP mass and mass difference between LSP and the lightest chargino state. Points obtained from seeded CMAES-ND scan and are color coded by the bino content (left) and higgsino content (right).}
	\label{fig:seeded_scan_m_Chi_1_Delta_m_Chi_1_Cha_1}
\end{figure}

\begin{figure}[H]
	\centering
	\includegraphics[width = 0.45\textwidth]{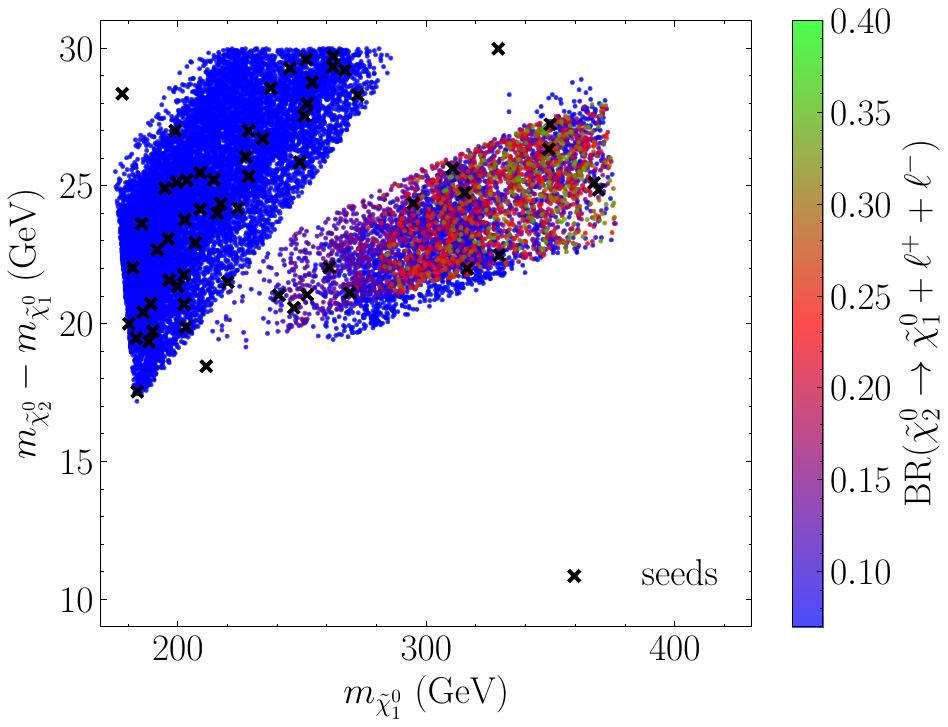}
    \includegraphics[width = 0.45\textwidth]{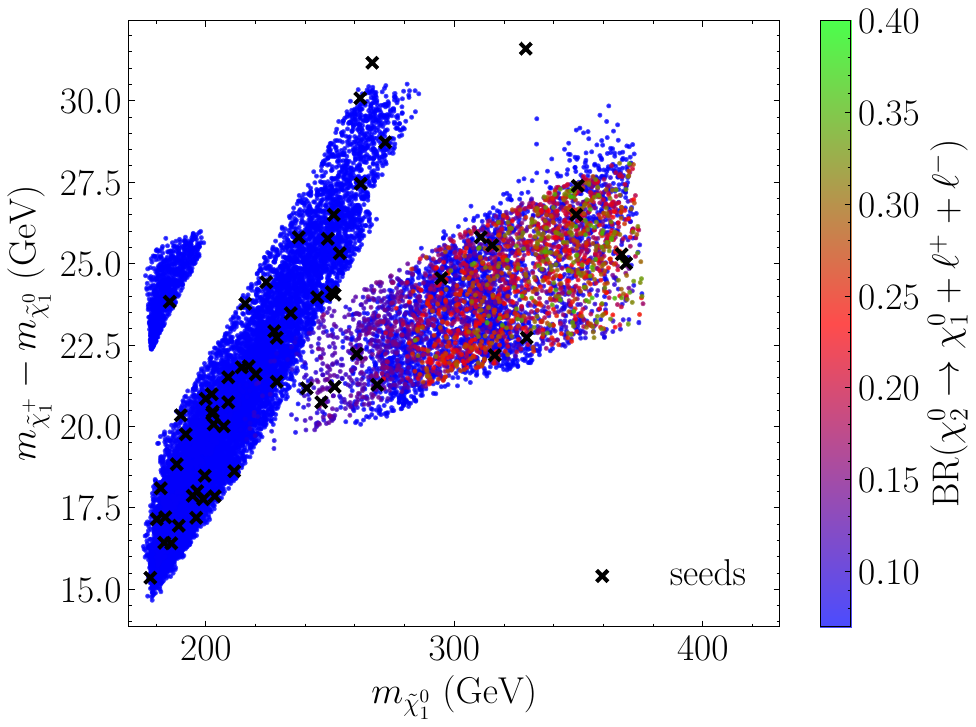}
	\caption{LSP mass and mass difference between LSP and the lightest neutralino state (left) and chargino state (right). Points obtained from seeded CMAES-ND scan and are color coded by the branching ratio of the lightest neutralino state to LSP + leptons. }
	\label{fig:seeded_scan_m_Chi_1_Delta_m_sum_BR}
\end{figure}

From the seeds drawn from the exploratory general scan above, we initialise runs using \texttt{"HBOS"} novelty reward. We further demand the seeds used in this scan to have $\text{BR}(\tilde{\chi}^{0}_2 \rightarrow \tilde{\chi}^{0}_1 + \ell^{+} + \ell^{-}) > 0.07$ ($\ell$ being $e$ or $\mu$) as this helps to explain the excesses even if the production is surpressed due to mixing effects in either the neutralino and/or chargino sector.  For this type of novelty reward, we choose $\{ m_{\tilde{\chi}^{0}_2}, m_{\tilde{\chi}^{0}_2} - m_{\tilde{\chi}^{0}_1}, m_{\tilde{\chi}^{\pm}_1} - m_{\tilde{\chi}^{0}_1}, \text{BR}(m_{\tilde{\chi}^{0}_2} \rightarrow m_{\tilde{\chi}^{0}_1} + \ell^{+} + \ell^{-}) \}$ to be the \emph{focus} observables, i.e. where the density of points will be computed in order to push the CMA-ES to acquire more diverse points in this subset of predictions. Results are shown in~\cref{fig:seeded_scan_m_Chi_1_Delta_m_Chi_1_Chi_2}, ~\cref{fig:seeded_scan_m_Chi_1_Delta_m_Chi_1_Cha_1} and~\cref{fig:seeded_scan_m_Chi_1_Delta_m_sum_BR}. Here we can see how HBOS ND has assisted CMA-ES to efficiently explore around the selected seeds and provide an expanded picture of the possible phenomenological predictions.
Specifically, solutions with high branching ratio of the second lightest neutralino state to LSP + leptons were discovered, as shown in~\cref{fig:seeded_scan_m_Chi_1_Delta_m_sum_BR}. These solutions were not found in the previous general scan without the novelty detection rewarding exploration over this quantity.

\subsubsection{Seeded Optimiser ND scan}

\begin{figure}[H]
	\centering
    \includegraphics[width = 0.45\textwidth]{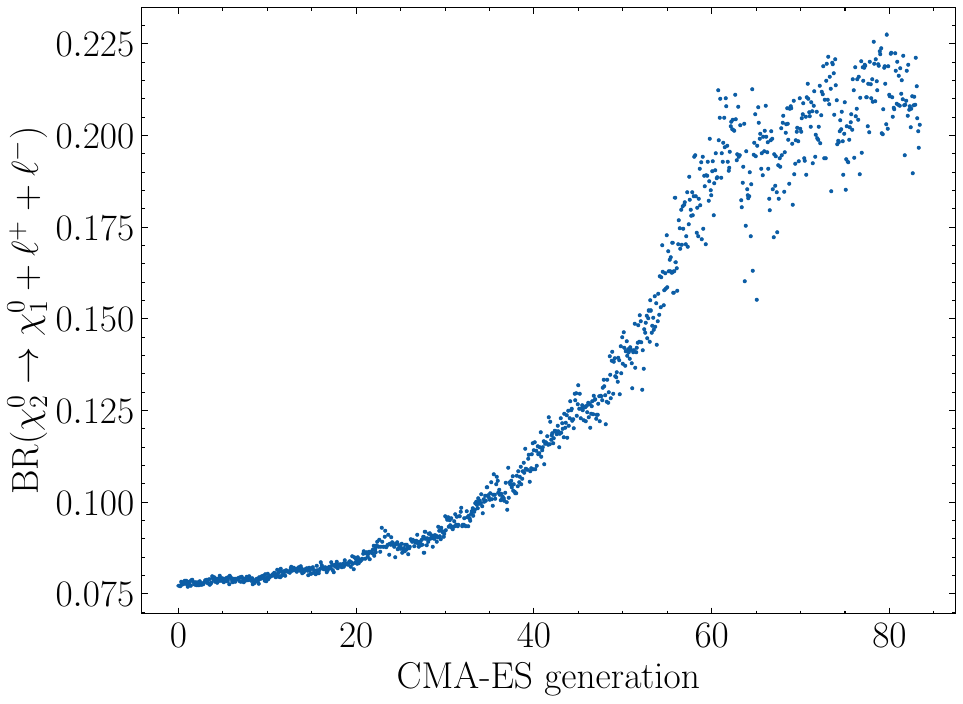}
	\caption{Evolution of branching ratios for neutralino decays to leptons in the Optimiser scan.
    }
	\label{fig:sum_BR_Chi_2_generation}
\end{figure}

The possibility for high branching ratios for the second lightest neutralino state decaying to the LSP+leptons can be further explored. Here we illustrate the functionality of the \texttt{"Optimiser"} novelty detection option, by using it in a seeded scan focused on maximising $\text{BR}(\tilde{\chi}^{0}_2 \rightarrow \tilde{\chi}^{0}_1 + \ell^{+} + \ell^{-})$. In~\cref{fig:sum_BR_Chi_2_generation} we see the evolution of the mean BR per iteration of CMA-ES. It clearly shows a successful steady growth in this observable after each generation, demonstrating the versatily of \verb|CMAES_ND| not only to map the viable regions of the parameter space shown above, but also to produce points with specific phenomenological characteristics that can be of interest to other downstream analyses.

\subsection{NMSSM scan with CMAES-ND}

In \cite{Ellwanger:2024vvs} a potential explanation for the electroweakino excesses \emph{and} the higgs-like excess at $95$ GeV was presented in the NMSSM. This involved a scenario where the soft lepton excess would be explained by higgsino-like neutralinos but the correct relic density of dark matter could be found by mixing with the bino and/or singlino. In particular, in order to mimic the higgsino signal at the LHC the lightest neutralino must be very close in mass to the lightest higgsino-like neutralino. However, that paper did not actually test the scenario's predictions for the signal.

In \cite{Araz:2025bww} this scenario was investigated using recasts of the relevant ATLAS soft lepton searches; parameter points were generated according to the scan region specified in the original paper, using a Markov Chain Monte Carlo scan in \BSMArt version 1.6 with a biased likelihood to find interesting regions of the parameter space. In this way,  representative points were found and analysed. Unfortunately, the scenario appeared to generally perform much worse than the simplified higgsino model, leading to exclusions rather than a best-fit region. However, it is likely that this was due at least in part to not selecting points with enough tuning of the neutralino masses: it was hard to find points with small mass splittings. 

Here we use the same parameter ranges as in \cite{Araz:2025bww} but employ a {\tt CMAES\_ND} scan. We choose the variables of interest to be: the masses of the lightest two Higgs bosons (which should be near $95$ and $125$ GeV respectively); mass splittings between the first two and second two neutralinos; the {\sc HiggsTools} {\sc HiggsBounds} signal ratio; the {\sc HiggsTools} {\sc HiggsSignals} $p$-value; the predicted signal rates $\mu_{\gamma\gamma}$ and $\mu_{bb}^{\rm LEP}$ which represent the rate of gluon fusion times decay to the corresponding channel compared to the SM rate for a Higgs boson of the same mass. These latter are the observables relevant for the putative $h_{95}$ excess (we also record the rate in the ditau channel, where there is possibly also an excess, but the model cannot accommodate the desired signal strength).
The input json file for the scan is provided in the examples repository on github, and we reproduce it in appendix \ref{APP:NMSSMJSON}.

\begin{figure}[H]
	\centering
    \includegraphics[width = 0.85\textwidth]{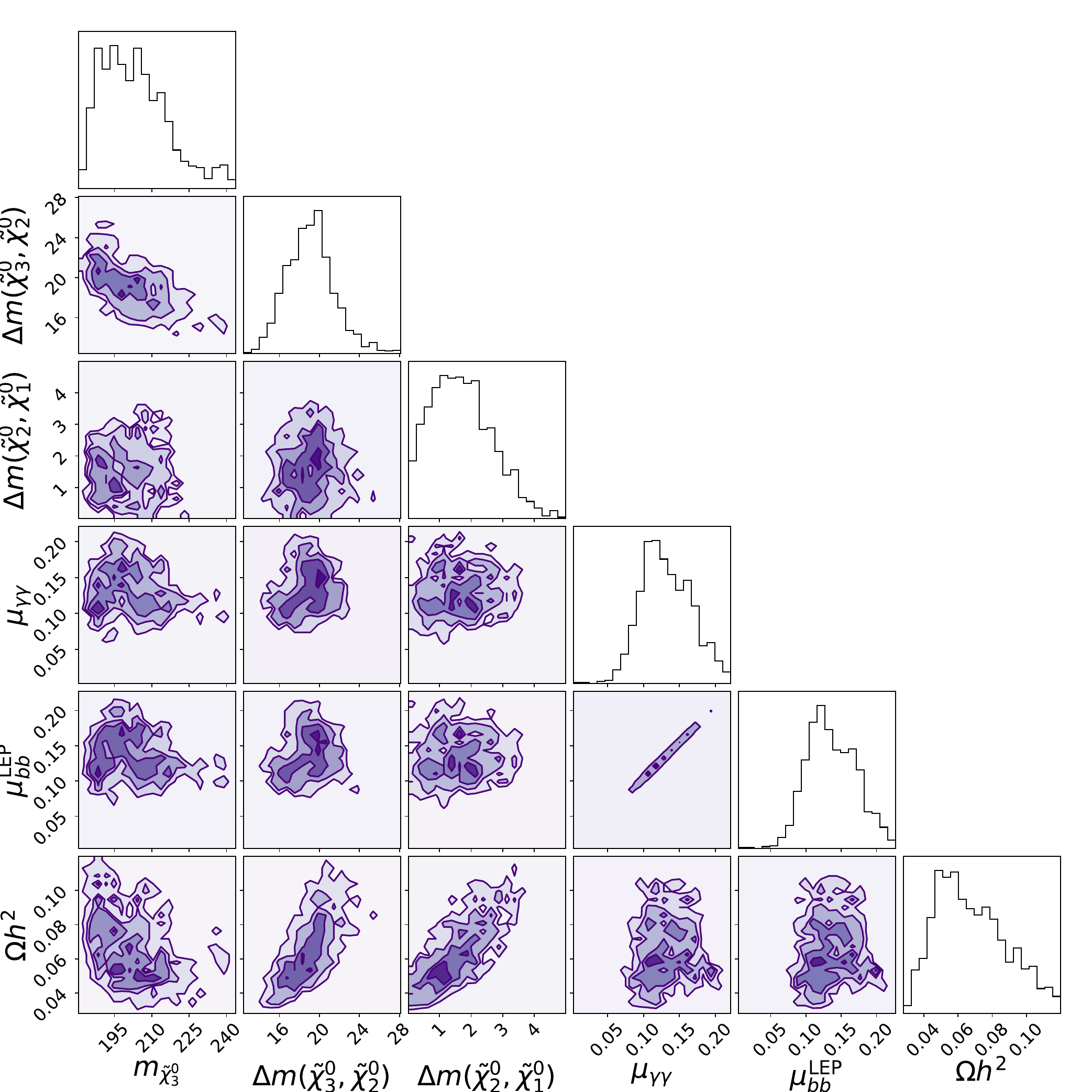}
	\caption{Corner plot of NMSSM observables}
	\label{fig:NMSSMcorner}
\end{figure}

The scan rapidly produced over a thousand valid parameter points. We present a corner plot of the interesting observables in figure \ref{fig:NMSSMcorner}. It is clear that the scan is rapidly able to find points with the desired properties; they can fit the $95$ GeV Higgs excess and also potentially satisfy the soft lepton excesses at the LHC. In future work we expect to use this dataset with the suite of recast analyses to fully test the scenario. Some pruning will be necessary because testing thousands of points with the recasting machinery is computationally prohibitive! However, passing them to {\sc HackAnalysis} or {\sc MadAnalysis} for this purpose can be handled with \BSMArt.

\section{Outlook}
\label{SEC:CONCLUSIONS}

We have presented version 2.0 of \BSMArt and some examples of its utility. As a versatile and powerful framework, we were able to implement versions of the CMA-ES algorithm that can be applied to scanning the parameter spaces of models of new physics, with or without novelty detection. The version without novelty detection is a highly efficient optimisation algorithm that can be used as a default scan where the aim is to find the ``best'' values of parameters for a model; this data can then be used to seed explorations of the model space around that point, for example using novelty detection. 

We gave two new examples relevant for soft lepton searches at the LHC where the new features of \BSMArt are useful. In the pMSSM, we showed how regions of the parameter space with enhanced dilepton branching ratios can be found; such points are useful as potential explanations of the excesses in themselves, but it would be very interesting to use these techniques in a full scan extending \cite{ATLAS:2024qmx,CMS-PAS-SUS-24-004,Constantin:2025bqp}.  In the context of the NMSSM, we showed how the new techniques can find regions of the parameter space that can explain both the 95 GeV excess and, potentially, the soft lepton excesses, in contrast to previous approaches. 

We hope that the latest version of \BSMArt is sufficiently easy to use and powerful to be adopted by people wishing to quickly use the associated tools; but also that it can be used by those interested in developing algorithms for parameter space exploration. The public \BSMArt examples repository is a step in that direction. There are many future directions that could be explored to exploit machine-learning techniques for different types of searches that we hope to return to in future work.

\section*{Acknowledgements}

M.D.G. is supported in part by Grant ANR-21-CE31-0013, Project DMwithLLPatLHC, from the French \textit{Agence Nationale de la Recherche} (ANR). M.D.G. and W.P. are funded in part by PROCOPE project 2025 / 52970ZL from Campus France and DAAD project-ID 57755908. We thank Asesh Datta for frequent feedback on the code;  Johannes Braathen, Luc Darmé, Martin Gabelmann, Wojciech Kotlarski and Taylor Murphy for feedback on and contributions to the code. M.C.R. is supported by the STFC under Grant No. ST/T001011/1. F.A.S. is also supported by FCT under the research grant with reference No. UI/BD/153105/2022. N.C. and F.A.S. are supported by FCT - Portugal under R\&{}D unit funding \href{https://doi.org/10.54499/UID/50007/2025}{UID/50007/2025}.


\appendix

\section{Scan information}
\label{app:scan}

\subsection{Parameters}

{\renewcommand{\arraystretch}{1.1}
\begin{table}[H]
	\makebox[\textwidth][c]{
		\begin{tabular}{|lr||lr|}
			\hline\hline
			Parameter                & Interval &
            \, Parameter                & Interval
			\\
			\hline
            $M_1$ & $[-4, 4]$ TeV & \, $|M_{\tilde{L}_1}|^{2}$ & $[0, 16] \text{ TeV}^{2}$  \\
            $M_2$ & $[-4, 4]$ TeV & \, $|M_{\tilde{e}_1}|^{2}$ & $[0, 16] \text{ TeV}^{2}$ \\
            $M_3$ & $[0, 10]$ TeV &  \, $|M_{\tilde{L}_3}|^{2}$ & $[0, 16] \text{ TeV}^{2}$ \\
            $\mu$ & $[-4, 4]$ TeV & \, $|M_{\tilde{e}_3}|^{2}$ & $[0, 16] \text{ TeV}^{2}$  \\
            \hline
            $A_t$ & $[-7, 7]$ TeV &  \, $|M_{\tilde{q}_1}|^{2}$ & $[0, 16] \text{ TeV}^{2}$ \\
            $A_b$ & $[-7, 7]$ TeV &  \, $|M_{\tilde{u}_1}|^{2}$ & $[0, 16] \text{ TeV}^{2}$ \\
            $A_{\tau}$ & $[-7, 7]$ TeV &  \, $|M_{\tilde{d}_1}|^{2}$ & $[0, 16] \text{ TeV}^{2}$ \\
            \cline{1-2}
             $|M_A|^{2}$ & $[0, 16] \text{ TeV}^{2}$ & \, $|M_{\tilde{q}_3}|^{2}$ & $[0, 16] \text{ TeV}^{2}$ \\
            $\tan \beta$ & $[2, 60]$ & \, $|M_{\tilde{u}_3}|^{2}$ & $[0, 16] \text{ TeV}^{2}$ \\ 
            & & \, $|M_{\tilde{d}_3}|^{2}$ & $[0, 16] \text{ TeV}^{2}$ \\ \hline\hline
        \end{tabular}
    }
    \caption{Ranges for the observable constraints considered for the seeded \BSMArt pMSSM CMAES-ND scan.}
    \label{tab:parameters}
\end{table}}

\subsection{Constraints\label{app:pmssm_constraints}}

Points with masses below 45 GeV are excluded by Z width. We also remove points excluded by LEP: $m_{\tilde{\chi}^{\pm}_1} < 91.9$ GeV for $\Delta m(\tilde{\chi}^{\pm}_1, \tilde{\chi}^{0}_1) < 3$ GeV and $m_{\tilde{\chi}^{\pm}_1} < 103$ GeV for $\Delta m(\tilde{\chi}^{\pm}_1, \tilde{\chi}^{0}_1) \geq 3$ GeV.

{\renewcommand{\arraystretch}{1.1}
\begin{table}[H]
	\makebox[\textwidth][c]{
		\begin{tabular}{lrr}
			\hline\hline
			Observable                & Allowed values & Scan type
			\\
			\hline
            $m_h$ & $[122, 128]$ GeV & Both\\
            $m_{\chi^{0}_2}$ & $[200, 400]$ GeV & Seeded\\
            $m_{\tilde{\chi}^{0}_2} - m_{\tilde{\chi}^{0}_1}$ & $[10, 30]$ GeV & Seeded\\
            $\text{BR}(\tilde{\chi}^{0}_2 \rightarrow \tilde{\chi}^{0}_1 + \ell^{+} + \ell^{-} )$ & $> 0.07$ & Seeded\\
            \hline
            $\Omega h^{2}$ & $[0.08, 1.14]$ & Both\\
            $\text{DM DD } p\text{-value}$ & $> 0.1$ & Both\\
            \hline
            $r$-value & $< 1$ & Both\\
            \hline\hline
        \end{tabular}
    }
    \caption{Ranges for the observable constraints considered for the  \BSMArt pMSSM CMAES-ND scans.}
    \label{tab:observables}
\end{table}}

\section{JSON input for the NMSSM CMAES-ND scan}
\label{APP:NMSSMJSON}

The input {\tt json} for the NMSSM scan of section \ref{SEC:PHYSICS} is:
\begin{minted}
[
frame=single,
framesep=2mm,
baselinestretch=1.2,
bgcolor=White,
fontsize=\footnotesize,
mathescape=true,
linenos,
]
{python}
{
    "Codes": {
        "SPheno": {
            "Command": "SPheno-4.0.5/bin/SPhenoNMSSM",
            "InputFile": "Scan_Spect.in.NMSSM",
            "OutputFile": "SPheno.spc.NMSSM",
            "Observables": {
                "mh1": {
                    "SLHA": ["MASS",[25]],
                    "SCALING": "CFUNCTION",
                    "RANGE": [92,98],
                    "VALID": [92,98]
                },
                "mh2": {
                    "SLHA": ["MASS",[35]],
                    "SCALING": "CFUNCTION",
                    "RANGE": [122,128],
                    "VALID": [122,128]
                },
                "MN1": {
                    "SLHA": ["MASS",[1000022]],
                    "SCALING": "OFF",
                    "MEAN": 133,
                    "VARIANCE": 5
                },
                "MN2": {
                    "SLHA": ["MASS",[1000023]],
                    "SCALING": "OFF"
                },
                "MN3": {
                    "SLHA": ["MASS",[1000025]],
                    "SCALING": "OFF"
                },
                "lambdaIn": {
                    "SLHA": ["NMSSMRUN",[1]],
                    "SCALING": "OFF"
                },
                "diff12": {
                    "FUNCTION": "abs(MN2)-abs(MN1)",
                    "SCALING": "CFUNCTION",
                    "RANGE": [0,5],
                    "VALID": [0,5]
                },
                "diff23": {
                    "FUNCTION": "abs(MN3)-abs(MN2)",
                    "SCALING": "CFUNCTION",
                    "RANGE": [10,30]
                }
            },
            "Run": "True"
        },
        "HiggsTools": {
            "HiggsBounds Dataset": "hbdataset-master",
            "HiggsSignals Dataset": "hsdataset-main",
            "Neutral Higgs": [25,35,45,36,46],
            "Charged Higgs": [37],
            "Signal Strengths": "True",
            "Observables": {
                "HBobsr": {
                    "SLHA": ["HIGGSTOOLS",[2]],
                    "SCALING": "CFUNCTION",
                    "RANGE": [0,1],
                    "VALID": [0,1]
                },
                "HSpval": {
                    "SLHA": ["HIGGSTOOLS",[13]],
                    "SCALING": "CFUNCTION",
                    "RANGE": [0.2, 1.0],
                    "VALID": [0.1,0]
                },
                "mugamgam": {
                    "SLHA": ["HIGGSTOOLS",[1003]],
                    "SCALING": "CFUNCTION",
                    "RANGE": [0.08,40]
                },
                "mubbLEP": {
                    "SLHA": ["HIGGSTOOLS",[1400]],
                    "SCALING": "CFUNCTION",
                    "RANGE": [0.05,2]
                },
                "mutautau": {
                    "SLHA": ["HIGGSTOOLS",[1006]],
                    "SCALING": "OFF",
                    "RANGE": [0.7,70]
                }
            },
            "Run": "True"
        },
        "MicrOmegas": {
            "Command": "micromegas_6.1.15/SARAH_NMSSM/MicrOmegas_v6.1_BSMArt_xsections",
            "InputFile": "SPheno.spc.NMSSM",
            "OutputFile": "omg.out",
            "Cross-sections": "True",
            "Run": "True",
            "LSP": [
                1000022
            ],
            "Observables": {
                "Oh2": {
                    "SLHA": ["DARKMATTER",[1]],
                    "SCALING": "CFUNCTION",
                    "RANGE": [0,0.13],
                    "LOG SCALING": "True"
                },
                "CDM1": {
                    "SLHA": ["DARKMATTER",[2]],
                    "SCALING": "OFF"
                },
                "DMpval": {
                    "SLHA": ["DARKMATTER",[401]],
                    "SCALING": "CFUNCTION",
                    "RANGE": [0.2,1.0],
                    "VALID": [0.1,0]
                }
            },
            "DD_Limits": "False"
        },
        "SModelS": {
            "Path": "smodels-main",
            "Cache": "smodels-main/CACHE",
            "QNUMBERS file": "QNUMBERS_NMSSM.slha",
            "Parameter file": "parameters.ini",
            "Observables": {
                "SModelSr": {
                    "SLHA": ["SMODELS",[1]],
                    "SCALING": "CFUNCTION",
                    "RANGE": [0,1]
                }
            },
            "Run": "True"
        }
    },
    "Setup": {
        "RunName": "NMSSM_ND",
        "Type": "CMAES_ND",
        "Cores": 12,
        "StoreEverything": "False",
        "StoreAllPoints": "False",
        "Store In Memory": "True",
        "Store Invalid": "True",
        "StoreSeparateFiles": "False",
        "CMAESPopulationSize": 24,
        "csv": "True",
        "Merge Results": "True",
        "Output File": "NMSSM_Output",
        "Spectrum File": "SPheno.spc.NMSSM",
        "Points": 10000,
        "Iterations": 30
    },
    "Variables": {
        "lamIn": {
            "RANGE": [0.013,0.019],
            "VARIANCE": 0.001
        },
        "kapIn": {
            "RANGE": [0.0058,0.0086],
            "VARIANCE": 0.001
        },
        "tbIn": {
            "RANGE": [3.61,10.9],
            "VARIANCE": 2
        },
        "AtIn": {
            "RANGE": [-5000,142],
            "VARIANCE": 500
        },
        "muIn": {
            "RANGE": [-244,-148],
            "VARIANCE": 10
        },
        "AlamIn": {
            "RANGE": [-5000,-1820],
            "VARIANCE": 500
        },
        "AkapIn": {
            "RANGE": [93,362],
            "VARIANCE": 50
        },
        "M1In": {
            "RANGE": [178,290],
            "VARIANCE": 10
        },
        "M2In": {
            "RANGE": [304,5000],
            "VARIANCE": 500
        },
        "M3In": {
            "RANGE": [739,5000],
            "VARIANCE": 500
        }
    },
    "Observables": {},
    "Fitting": {
        "Variables": {
            "MQ3In": {
                "RANGE": [73984,25000000]
            }
        },
        "Observables": {
            "mh2": {
                "TARGET": 125,
                "VARIANCE": 0.5
            }
        },
        "Options": {
            "eps": 0.1,
            "ftol": 1.0,
            "disp": "False",
            "Overloads": [["SPHENOINPUT",[11],0]]
        }
    },
    "Plotting": {
        "Strategy": "csv",
        "Plots": {
            "mh1_vs_mh2": {
                "Labels": [
                    "$m_{h1}$ (GeV)",
                    "$m_{h2}$ (GeV)"
                ],
                "Values": [
                    "mh1",
                    "mh2"
                ],
                "Ranges": [[60,120],[118,128]]
            }
        }
    }
}
\end{minted}

\bibliographystyle{utphys}
\bibliography{refs}

\end{document}